\def\ispreprint{1}  

\if \ispreprint1
\documentclass[preprint,authoryear,5p]{elsarticle}  
\else
\documentclass[preprint,authoryear,12pt]{elsarticle}  
\fi

\usepackage[table]{xcolor}
\usepackage[utf8]{inputenc}
\usepackage{dsfont}
\usepackage{graphicx}
\usepackage{graphicx}
\usepackage{gensymb}  
\usepackage{tabularx}
\usepackage{amssymb}
\usepackage{amsmath}
\usepackage{hyperref}
\usepackage{multirow}
\usepackage{placeins}
\usepackage{xparse}
\usepackage{natbib} 
\usepackage{lineno} 
\usepackage{setspace}
\usepackage{caption}

\hypersetup{draft}

\begin{document}

\title{Propagation and attenuation of pulses driven by low velocity normal impacts in granular media} 

\author[a1]{A. C. Quillen\corref{cor1}}
\ead{alice.quillen@rochester.edu}

\author[a1]{Max Neiderbach}
\ead{mneiderb@u.rochester.edu}

\author[a1]{Bingcheng Suo}
\ead{bsuo@u.rochester.edu}

\author[a1]{Juliana South}
\ead{jsouth@u.rochester.edu}

\author[a1]{Esteban Wright}
\ead{ewrig15@ur.rochester.edu}

\author[a1]{Nathan Skerrett}
\ead{nskerret@u.rochester.edu}

\author[a2]{Paul S\'anchez}
\ead{diego.sanchez-lana@colorado.edu}

\author[a3]{Fernando David C\'u\~nez}
\ead{fcunezbe@ur.rochester.edu}

\author[a4]{Peter Miklavcic} 
\ead{pmiklavc@ur.rochester.edu} 

\author[a4]{Hesam Askari}
\ead{askari@rochester.edu}


\address[a1]{Department of Physics and Astronomy, University of Rochester, Rochester, NY 14627, USA}

\address[a2]{Colorado Center for Astrodynamics Research, The University of Colorado Boulder, 3775 Discovery Drive, 429 UCB - CCAR, Boulder, CO 80303, USA}

\address[a3]{Department of Earth and Environmental Science, University of Rochester, Rochester, NY 14627, USA}

\address[a4]{Department of Mechanical Engineering, University of Rochester, Rochester, NY 14627, USA}


\cortext[cor1]{Corresponding author}


\begin{abstract}
We carry out experiments of low velocity normal impacts into granular materials that fill an approximately cylindrical 42 litre tub.   Motions in the granular medium are tracked with an array of 7 embedded accelerometers.  Longitudinal pulses excited by the impact attenuate and their shapes broaden and become smoother as a function of travel distance from the site of impact.   Pulse propagation is not spherically symmetric about the site of impact. Peak amplitudes are about twice as large for the pulse propagating downward than at 45 degrees from vertical. An advection-diffusion model is used to estimate the dependence of pulse properties as a function of travel distance from the site of impact.   The power law forms for pulse peak pressure, velocity and seismic energy depend on distance from impact to a power of -2.5 and this rapid decay is approximately consistent with our experimental measurements.  Our experiments support a seismic jolt model, giving rapid attenuation of impact generated seismic energy into rubble asteroids, rather than a reverberation model, where seismic energy slowly decays.   We apply our diffusive model to estimate physical properties of the seismic pulse that will be excited by the forthcoming DART mission impact onto the secondary, Dimorphos, of the asteroid binary (65803) Didymos system.   We estimate that the pulse peak acceleration will exceed the surface gravity as it travels through the asteroid. 


\end{abstract}

\maketitle

\if\ispreprint1
\else 
\linenumbers 
\fi 

\section{Introduction}
\label{sec:intro}

Apollo-class Near-Earth Asteroid binary (65803) Didymos is the target of the international collaboration known as AIDA (abbreviation for Asteroid Impact \& Deflection Assessment) that supports the development and data interpretation of the NASA's Double Asteroid Redirection Test (DART) mission \citep{Cheng_2018_DART,Rivkin_2021} and the European Space Agency's Hera mission \citep{Michel_2022}.
Goals of these missions to the potentially hazardous binary asteroid Didymos include measuring the momentum transfer efficiency and resulting deflection from a hyper-velocity asteroid impact. 
DART will be the first high-speed impact experiment on an asteroid at a scale relevant for planetary defense.  Imaging during the impact will be carried out by the accompanying 6U CubeSat named the Light Italian CubeSat for Imaging of Asteroids (also known as LICIACube; \citealt{Dotto_2021}). 

A high velocity impact compresses the target material to high pressures, launching a shock wave which causes vaporization, melting,
fragmentation, plastic deformation, and formation of a crater \citep{melosh89}.  
The expanding shock wave propagates outward from the impact site and attenuates as it propagates.  When the velocity of the shock drops below
a certain threshold, it continues to propagate as an elastic wave \citep{holsapple93}. 
Most Near Earth Asteroids (NEAs) are expected to be comprised of rubble \citep{walsh18} which 
complicates predicting the strength and attenuation rate of impact induced seismic energy \citep{mcgarr69}.

There are two views on how 
impact generated seismic energy propagates within rubble asteroids   
\citep{cintala78,cheng02,richardson04,thomas05,Yamada_2016}.
The rapidly attenuated seismic pulse or `jolt' model \citep{nolan92,greenberg94,greenberg96,Nolan_2001,thomas05} 
is consistent with strong attenuation in dry laboratory granular materials at kHz frequencies \citep{Hostler_2005,Odonovan_2016}. However, the jolt model qualitatively differs from
the slowly attenuating seismic reverberation model \citep{cintala78,cheng02,richardson04,richardson05,Yamada_2016}, that is supported
by measurements of slow seismic attenuation rates in lunar regolith \citep{dainty74,toksoz74,nakamura76}.
While both impact-induced seismic jolt
and reverberation can cause crater erasure, crater rim degradation and resurfacing \citep{veverka01,Nolan_2001,richardson04,richardson05,thomas05,asphaug08,Yamada_2016},
size segregation induced by the Brazil-nut effect could depend on sustained vibrations or reverberation
(e.g., \citealt{miyamoto07,tancredi12,matsumura14,tancredi15,perera16,maurel17,toshihiro18}), though a single seismic pressure pulse can also leave boulders on the surface \citep{wright19} via ballistic sorting \citep{shinbrot17}.

Pulse propagation in granular media is nonlinear, sensitive to pulse duration and amplitude, ambient or hydrostatic pressure and the nature of contacts in the granular medium (e.g., \citealt{Goddard_1990,Liu_1992,Jia_1999,Johnson_2000,Hostler_2005,Bi_2011,Gomez_2012}).    At low pulse amplitude, the pulse propagation speed along the fastest travel path along one chain of contacts can differ from the propagation speed of a pulse peak \citep{Liu_1992,Owens_2011}.   In some regimes, a pulse can propagate as if it were a coherent elastic wave or sound wave \citep{Geng_2003,Somfai_2005,vandenWildenberg_2013,Santibanez_2016},  while in other regimes, diffusive, dispersive and anisotropic behavior is predicted or observed \citep{DaSilva_2000,Otto_2003,Jia_2004,Luding_2005,Hostler_2005}. 
Hertzian (or Hertz-Mindlin) contact theory underlies estimates for the nature of sound propagation 
in granular media 
(e.g., \citealt{Gomez_2012,vandenWildenberg_2013}) and development of a continuum model, denoted an effective medium theory  (e.g., \citealt{Goddard_1990,Johnson_2000}). 
In granular systems, broadening and attenuation of seismic or acoustic pulses may be intrinsically related  \citep{Jia_2004,Langlois_2015,Odonovan_2016,Zhai_2020}.


The difficulty of predicting impact induced seismicity in granular systems has motivated experiments
that measure the response of granular materials to impacts.  \citet{mcgarr69} conducted impact experiments on epoxy-bonded sand and on unconsolidated or loose sand at impact velocities of 0.8 to 7 km/s.
They measured accelerations using accelerometers placed on the substrate or target surface.   
The signals
from the bonded sand experiment were sinusoidal, showing many periods of oscillation, however,
the accelerations in the unconsolidated sand resembled a single period of a sine wave (see their Figure 3).  \citet{yasui15} carried out a series of intermediate impact velocity experiments,
with impact velocity approximately 100 m/s 
into spherical glass beads with diameter 180--250 $\mu$m.  
\citet{Matsue_2020} extended this work with impact experiments into quartz sand at higher velocities ranging from 200 m/s to 7 km/s.   The accelerometer signals from these two sets of experiments resembled those
seen by \citet{mcgarr69} in unconsolidated sand, confirming that a single seismic pulse tends
to be excited in a granular substrate by an impact.  
\citet{yasui15} and \citet{Matsue_2020} 
showed that the strength of the peak acceleration  $g_{max}$ 
in the impact generated seismic signal decayed as a function of distance $r$ from impact site with
a power law function $g_{max} \propto r^\gamma$ with exponent  $\gamma = -2.21 \pm 0.12 $ \citep{yasui15} and $-3.12 \pm 0.10$ \citep{Matsue_2020}.  
 
In this paper, using an array of 7 accelerometers,
we examine how impact excited pulses travel in granular media.
Our impacts are low velocity (a few m/s) normal impacts 
of small spherical projectiles into 41.6 litres 
of sand or millet contained
in an approximately cylindrical tub.  Our experiments are at lower velocity than those
of \cite{yasui15} and \citet{Matsue_2020} and we go beyond
theses studies by comparing  the impact generated seismic pulses in different
granular substrates.  We use our accelerometer array to study variations in the signal shapes
as a function of depth and distance from impact.  
Our experiments are described
in Section \ref{sec:methods}.  In Section \ref{sec:exp} we examine the peak values and durations of pulses 
and how the pulses propagate through the granular medium.   
Motivated by the rapid attenuation and smoothing and broadening of the pulse shapes, in Section \ref{sec:diff} we use a diffusive attenuation model 
to study the dependence of pulse duration, peak amplitude
and other quantities on distance from impact site.  
The DART impact gives unprecedented opportunity to probe how a seismic wave generated by a hypervelocity impact is transmitted through asteroid granular material
and affects the surface.
In Section \ref{sec:DART} we discuss implications of our experiments and model
for the forthcoming DART mission impact.
A summary and discussion follow in Section \ref{sec:sum}.

\begin{table*}[htbp] 
\if \ispreprint1
\else\tiny
\fi
\caption{Accelerometer placement coordinates $(R, \theta, z)$} \label{tab:template}
\begin{tabular}{lllllllllll}
\hline
   & A          & B           & C              & D             & E            & F           & G \\
 \hline 
L5    & (5,0,-5) & (8,0,-5) & (11,0,-5) & (14,0,-5) & (17,0,-5) & (20,0,-5) & (23,0,-5) \\
L10 \!\!  & (2,0,-10)  & (5,0,-10) & (8,0,-10) & (11,0,-10) & (14,0,-10) & (17,0,-10) & (20,0,-10)  \\
L15 \!\! & (2,0,-15)  & (5,0,-15) & (8,0,-15) & (11,0,-15) & (14,0,-15) & (17,0,-15) & (20,0,-15)  \\
R5 &(5.5,-60,-4)&(5.5,-40, -7)&(5.5,-20, -10)&(5.5,0,-13) & (5.5,20,-16)& (5.5,40,-19) & (5.5,60,-22)  \\
R10 \!\! &(10.5,-45,-4)&(10.5,-30, -7)&(10.5,-15, -10)&(10.5,0,-13) & (10.5,15,-16)& (10.5,30,-19) & (10.5,45,-22)  \\
R15 \!\! &(15.5,-45,-4)&(15.5,-30, -7)&(15.5,-15, -10)&(15.5,0,-13) & (15.5,15,-16)& (15.5,30,-19) & (15.5,45,-22)  \\
\hline 
\end{tabular}
\\ \par {\footnotesize
\begin{singlespace}
Notes: Each row gives a set or template of accelerometer coordinates.
These are cylindrical coordinates $(R,  \theta$,z), for each accelerometer with $R, z$ in cm and $\theta$ in degrees.
The impact point is near the origin and on the surface.  
The column heads refer to the oscilloscope channels and 
the eighth or H oscilloscope channel is used to trigger
data recording with the IR break-beam sensor.   
The coordinate locations refer to the position of the accelerometer ADXL335 integrated circuit.
The angles are crudely estimated and should not affect the measurements because of the cylindrical symmetry of our experiment.\end{singlespace} }
\end{table*}

\begin{figure}[htpb] \centering
\includegraphics[width=3 truein, trim = 0 0 0 0,clip]{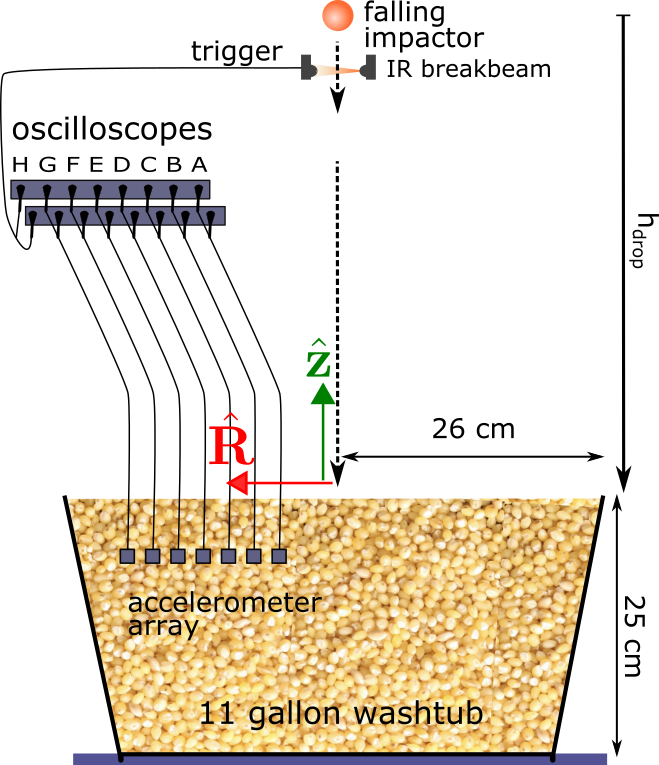}
\caption{A side view illustration of the experiment, including the washtub, granular material, spherical impactor
and accelerometer array.  The unit vectors $\hat {\bf R}, \hat{\bf z}$ are cylindrical coordinate directions with respect to the impact point.
\label{fig:setup}}
\end{figure}

\begin{figure}[htbp] \centering  
\if \ispreprint1
\includegraphics[width = 3.2 truein, trim = 70 50 10 20, clip]{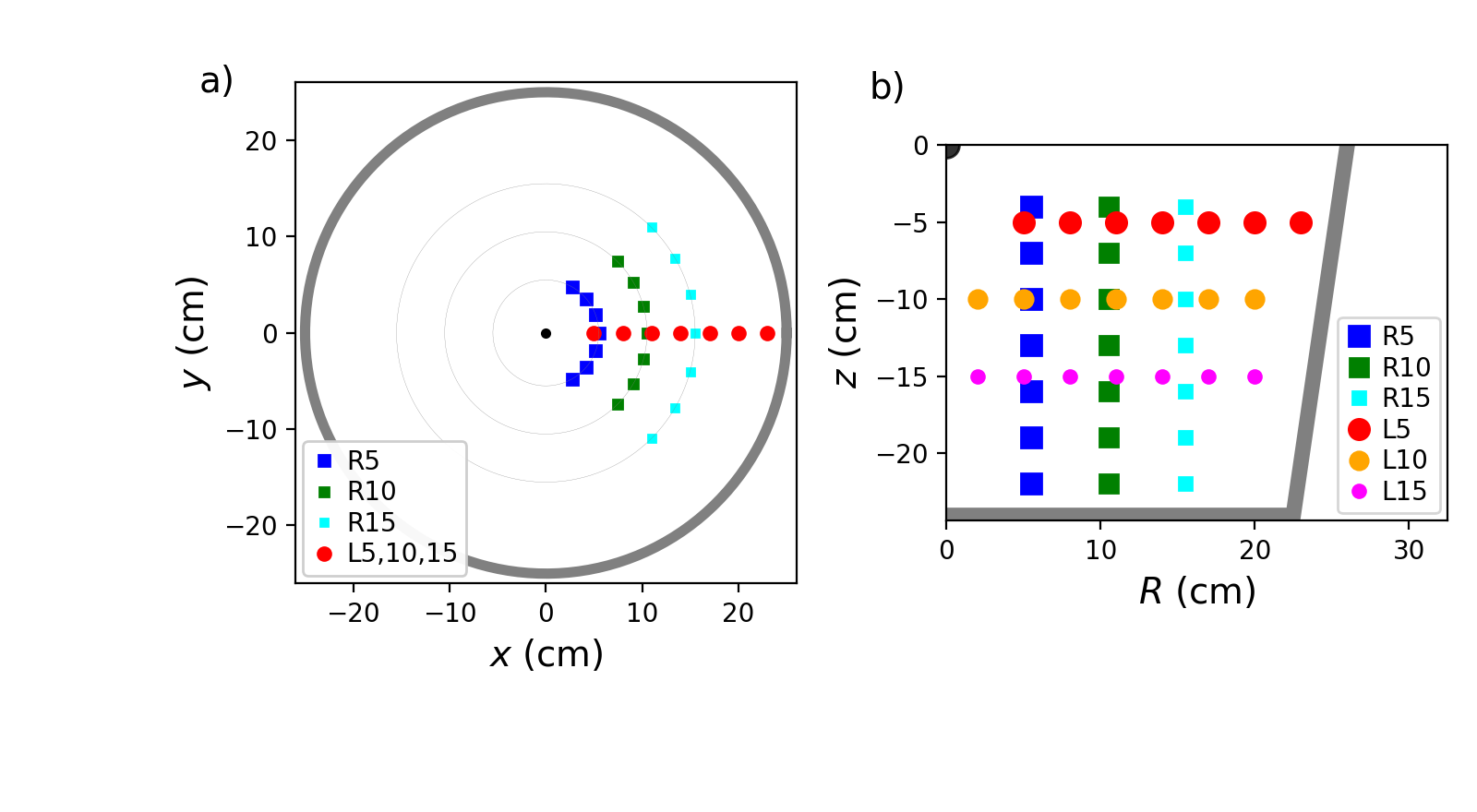}
\else
\includegraphics[width = 4.0 truein, trim = 70 50 10 20, clip]{templates.png}
\fi
\caption{Illustration of the accelerometer placements. 
a) A top down view of the accelerometer positions.  The thick grey circle shows
the outline of the rim of the tub.  
Each coordinate template consists of 7 accelerometer positions and is shown with a different color and size marker. The coordinate templates are listed in Table \ref{tab:template}.
The point of impact is shown with a black dot at the origin. 
b)  Similar to a) except showing a side view and plotting cylindrical radius $R$ vs $z$. 
The thick grey line shows the outline of the tub. 
The substrate surface is at $z=0$. 
  \label{fig:templates}
}
\end{figure}

\begin{figure*}[htbp] \centering 
\includegraphics[width = 6.2 truein, trim = 5 10 0 0, clip]{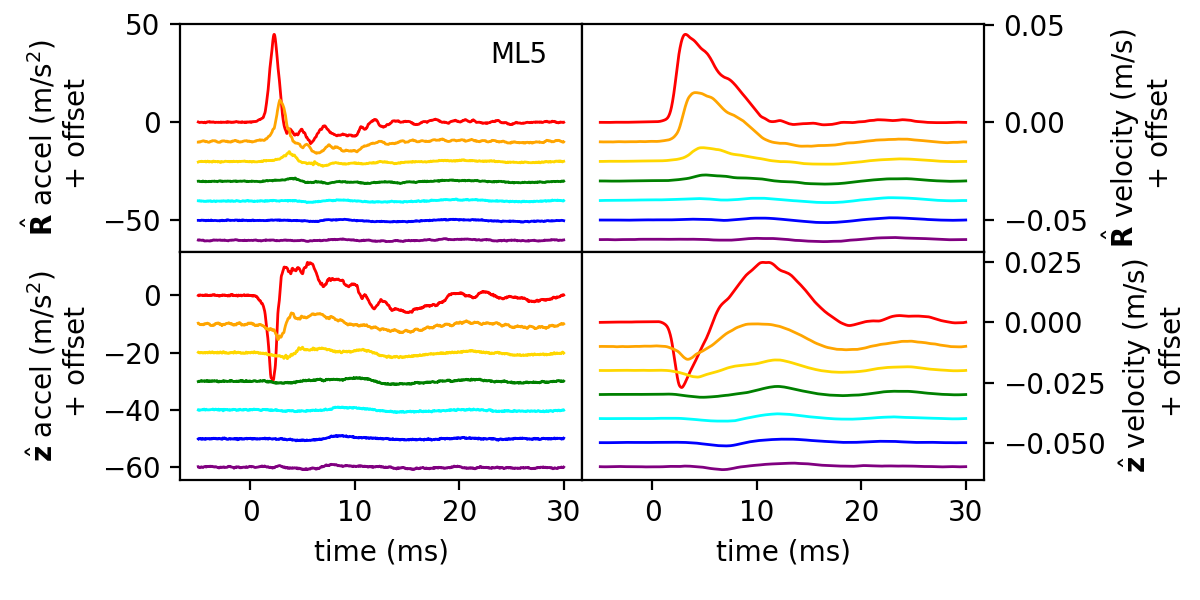}
\caption{ 
Accelerations measured in a 7 channel accelerometer array as a function of time following an impact
for experiment ML5 into millet. 
The accelerometers are at the same depth and in a line with increasing distance from
the impact site (the L5 coordinate template in Table \ref{tab:template}).
The top panels show the radial $\hat {\bf R}$ (cylindrical radius) components of acceleration and velocity
and the bottom two panels show the vertical, $\hat{\bf z}$, components.  
The left two panels show acceleration and the right two panels show velocities.
Signals from 7 accelerometers are plotted in each panel with a vertical offset between 
channels that is 10 m/s${^2}$ for accelerations and 0.01 m/s for velocities.   
In all panels the plots are shown in order of
distance from the impact site with the topmost signal from 
 the accelerometer nearest the impact site.
Each accelerometer is shown with a different color line.
\label{fig:ML5}} 
\end{figure*}

\begin{figure*}[htbp] \centering 
\includegraphics[width = 6.2 truein, trim = 5 10 0 0, clip]{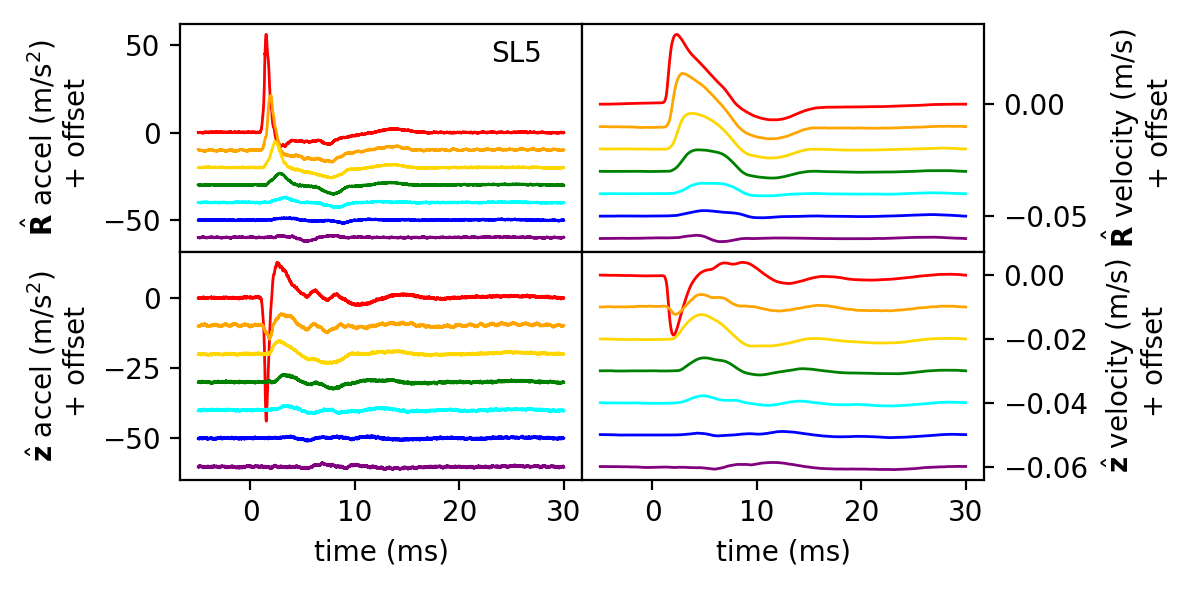}
\caption{ 
Similar to Figure \ref{fig:ML5} except showing the SL5 experiment with the same coordinate template but into fine sand.  The pulse is narrower in this experiment than the similar ML5 experiment into millet.
\label{fig:SL5}} 
\end{figure*}

\section{Experimental Methods}
\label{sec:methods}

Our granular substrate is held in a large 41.6 litre (11 gallon) washtub that is 25 cm deep and 52 cm in diameter at the top.  
The tub base is a circle with a diameter of 45 cm.  
The tub is filled with millet or sand.  
For an illustration of our experimental setup see Figure \ref{fig:setup}.

%

\subsection{The array of accelerometers}

We measure the impulse caused by an impact using
an array of 7 accelerometers that are embedded within the granular substrate. 
The accelerometers are 5V-ready analog break-out boards 
by Adafruit 
which house the $\pm 3 g$ triple-axis accelerometer ADXL335 Analog devices 
integrated circuit. 
The dimensions of the accelerometer printed circuit boards (PCBs) are 19 mm $\times$ 19 mm $\times$ 3 mm. 
The ADXL335 specifications describe its $z$ axis as that parallel to the narrowest dimension of the integrated circuit (and perpendicular to the PCB) and 
$x,y$ axes in the direction parallel to its two longer dimensions (and in the plane of the PCB).
We removed three on-board filter capacitors from the PCB to increase the output bandwidth upper limit from 50 Hz to 1600 Hz on $x$- and $y$-axes and to 550 Hz on the $z$-axis.  
We only use the $x$ and $y$ axis accelerometer signals because they have a higher bandwidth upper limit. The bandpass upper limits are frequencies
at which the signal amplitude is reduced by 3 dB (the amplitude drops by a factor of 0.5) and approximately equal to the cutoff frequency of a low pass filter. The 1600 Hz bandwidth upper limit correspond to a half period of 0.3 ms which is shorter than the width of the acceleration pulses seen in our experiments.
So that the accelerometers are free to move within the media, we used fine and flexible 36 AWG gauge wire to power the accelerometers and connect their signal outputs.  

Each accelerometer was individually calibrated in each axis by placing it on a level surface, taking a measurement, rotating it by $180^\circ$ around a horizontal axis and then repeating the measurement.   The difference between the two measured voltages are equivalent to 2 $g$, giving a calibration factor from volts to m/s$^2$.
The calibration factors are within $\pm$  0.002 V per m/s$^2$ of the mean value 
of all 7 accelerometers (that is $0.032\ V$ per m/s$^2$).  
We also recorded the DC voltage of each axis when aligned horizontally.
These were used to check the accelerometer orientations.

The output signals of the $x$ and $y$-axis outputs of the accelerometers were recorded with 
 8-channel digital oscilloscopes (Picoscope model 4824A) with a sampling rate of 100 kHz.
We use a range for the oscilloscope channels  of 
$\pm 5 V$  for each accelerometer 
giving 0.07 m/s$^2$ of precision in the acceleration measurements.  

\subsection{Triggering}

The impactors are solid spherical objects like a rubber ball or glass marble, that are released from a height of $h_{drop}\sim 1$ m above the substrate surface.  A small iron washer is glued to the ball.  
The ball is held in place with a solenoid that releases the ball when the solenoid is disconnected from DC power.  
As the ball falls, it passes between the transmitter and receiver of an IR break-beam sensor pair and this
triggers recording of the accelerometer signals.  
Rubber and glass are not perfectly opaque to infrared light and the projectiles surfaces can reflect light.
To improve the accuracy of the trigger timing, we painted the projectile surfaces matte black with common oil paint. 
The two oscilloscopes are triggered to record 14 channels of accelerometer data at the same time.  
One oscilloscope is used to record the $x$-axis accelerometer outputs and the other is used to record the $y$-axis outputs. 
In each oscilloscope 
channels 1 to 7 (also denoted channels A through G) are used to record accelerometer data and the 8-th, or H-th, channel is used to trigger the recordings.  

\subsection{Accelerometer placement}

Prior to each impact experiment we rake the substrate a few times, with a rake
that has prongs that are about 10 cm long
and are equally spaced by a cm.  The rake is also used to level the surface.
After raking, the accelerometers are embedded in the medium. 
We orient the accelerometers so that their $+x$ axes point away from the impact site and their $+y$ axes point vertically up. 
To ensure that the accelerometers are correctly spaced, at the desired depth, and correctly
oriented, we individually placed each accelerometer in the medium.  
For the shallow depths (less than 5 cm), we 
used a pair of long tweezers.  For the deeper locations we used  
a PVC tube with filed slots to hold the accelerometer board while we slowly pushed it into the granular medium.
The DC voltage levels of each accelerometer were monitored in both axes during placement to ensure that their orientations are approximately consistent with the desired orientation.  
We compared the DC voltage levels of the accelerometer signals prior to impact to the calibration values and find that the accelerometers, once embedded, are typically within $10^\circ$ 
of the desired orientation.

We adopt a coordinate system with origin at the impact site. In Cartesian coordinates
$(x,y,z)$ the vertical and $z$ coordinate is negative below the horizontal substrate  surface.
It is convenient to use a cylindrical coordinate system $(R, \theta, z)$ to describe accelerometer positions. 
Radius $R$ is the radial distance to the vertical line going through the impact point. 
The $z$ coordinate gives height from the initially flat granular surface.   
Directions for motions are described in terms of 
a radial unit vector in cylindrical coordinates $\hat {\bf R} = \frac{1}{R} (x,y,0)$, 
where on the right we write the vector in Cartesian coordinates, and
a vertical unit vector $\hat {\bf z} = (0,0,1)$ (in either cylindrical coordinates or Cartesian coordinates).
Since our accelerometers are below the surface
we describe their vertical position in terms of a depth which is $-z$.
Acceleration in the $+x$-axis direction with respect to the accelerometer chip 
gives measurements of acceleration in the radial $+ \hat {\bf R}$ 
direction.  Acceleration in the $+y$-axis direction with respect to the accelerometer chip gives measurements of acceleration in the vertical cylindrical coordinate or $+ \hat {\bf z}$ direction.
It is also convenient to use a spherical coordinate system to describe subsurface motions.
The radius from the impact site $r =  \sqrt{x^2 + y^2 + z^2} = \sqrt{R^2 + z^2}$.
The radial direction from the impact site is unit vector $\hat {\bf r} = \frac{1}{r}(x,y,z)$,  where
on the right we have the vector in Cartesian coordinates.

The 7 accelerometers were placed in the granular medium in sets of locations which we refer to as coordinate templates.
The cylindrical coordinates of each accelerometer are listed in Table \ref{tab:template} for 
each coordinate template.  Templates denoted with letter $L$ have accelerometers
placed in a line, at different radii and all at the same depth. 
Templates denoted with letter $R$ have accelerometers all at the same cylindrical radius, but at different depths and polar angles.

\begin{table}[htbp] 
\caption{Properties of Spherical Projectiles \label{tab:imp}}
\if \ispreprint1
\begin{tabular}{p{32mm}p{19mm}p{10mm}p{13mm}lll}
\else
\begin{tabular}{p{45mm}p{25mm}p{15mm}p{18mm}lll}
\fi 
\hline
                       & & Rubber & Glass \\
                       & & Ball  & Marble \\
\hline
Radius & $R_p$ (cm)                   & 1.83 & 1.41 \\
Mass  & $m_p$ (g)                        & 27.3 & 30.9 \\
Density & $\rho_p$ (g cm$^{-3}$)  &  1.06 &  2.63  \\
Young's modulus & $E_p$  (GPa) &  $ \sim\!$ 0.01 & $\sim $  50  \\
Sound prop. velocity & $c_p$  (m/s)   & $\sim\!$ 100 & $\sim $ 4400  \\
Sound prop. time & $t_{D,prop}$ (ms) & $\sim\!$ 0.7 & $\sim $ 0.012 \\
\hline
\end{tabular}
\\ \par { \footnotesize
\begin{singlespace}
Notes: 
We measured the mass and radius of each projectile. 
The sound velocity is an estimate for that within the projectile and the sound propagation
time is the time for a sound wave to traverse the object's diameter twice. 
\end{singlespace}
}
\end{table}

\subsection{Properties of granular media and impactors}

Spherical impactor masses $m_p$ and radii $R_p$ are listed in Table \ref{tab:imp}.  Impactors (which we also call projectiles) were chosen so that their density $\rho_p$ approximately matches that of the granular substrate.   For the experiments in sand we used a green glass marble and for the experiments into millet we used a colorful rubber ball.
This reduces sensitivity to the substrate to projectile density ratio that is present in crater scaling laws \citep{holsapple93}.
For the projectiles we list rough estimates for their Young's modulus $E_p$,  sound propagation speed $c_p \approx \sqrt{E_p/\rho_p}$, and time for a sound wave to propagate back and forth across the projectile $t_{D,prop} = 4 R_p/c_p$.

The properties of the granular media are listed in Table \ref{tab:granular}. 
Millet has the advantage that it is low density and this facilitates placing accelerometers deep within the medium.
Sand has the advantage that its grain properties are similar to rocky materials
that might be present on planetary surfaces. 
The millet is white proso millet marketed as birdseed.
The sand is the fine light playground sand described   
by \citet{Wright_2022}, and was passed through a sieve to remove
particles greater than 0.5 mm in width.  Both substrates are inexpensive. 
The procedures for measuring bulk density $\rho_s$, grain density $\rho_g$, porosity  $\phi$, 
 angle of repose $\theta_r$ and static friction coefficient $\mu_s$, are described by \citet{Wright_2022}. 
We adopt the Young's modulus $E_g\approx 100$ MPa for millet grain material measured  by \citet{millet_2015} using a Hertzian contact model. Sound speeds within the grains are estimated with 
$c_g \approx\sqrt{E_g/\rho_g}$. 
For sand grains we use the Young's modulus of $E_g \sim 10$ GPa which gives
a seismic p-wave velocity of about $c_g \sim 2$ km/s, similar to that in soft sandstones.
We list typical grain axis lengths for the millet in Table \ref{tab:granular}.
The millet seed sizes are fairly similar but they are not spherical. The fine sand has a wide grain size distribution, with a FWHM in the major axis distribution about 0.4 times the mean value 
(see Figure 3d by \citealt{Wright_2022}).

\begin{table}[htbp] 
\caption{Properties of Granular Media}
\label{tab:granular}
\begin{tabular}{lllllll}
\hline
           &          &  Millet   & Sand  \\
\hline 
Grain density  &  $\rho_g\!$\! (g cm$^{-3}$)  &  1.22      & 2.5   \\ 
Bulk density   &  $\rho_s\!$\! (g cm$^{-3}$)  &    0.75     &   1.5  \\
Porosity & $\phi$                                     &   0.39          & 0.42  \\
Angle of repose & $\theta_r$                  &  24\degree  & 36\degree  \\
Static friction coef.  & $\mu_s$         &   0.45         & 0.8  \\
Grain lengths or diam.  & $d_g$ (mm)     & $3.5\! \times\! 3\! \times\! 2 \! \! \!$ & $\sim\!0.25$   \\
Grain elastic modulus   &   $E_g$ (GPa)      & $\sim$ 0.1   & $\sim$ 10 \\
Grain sound speed     &    $c_g$  (m/s)      & $\sim$ 300  & $\sim$ 2000 \\
\hline
\end{tabular} \\  { \footnotesize
\par \begin{singlespace}
Notes: The coefficient of static friction for the granular material is computed from its angle of repose $\mu_s = {\rm tan}(\theta_r)$.
Our millet is white proso millet.  
The sand is the same as the fine light playground sand described by \citet{Wright_2022}.
The fine sand mean major axis length is 0.32 mm and the mean middle axis length
is 0.24 mm \citep{Wright_2022}.  The millet grain mass is about 6.5 mg and the mass
of a sand grain is about 0.3 mg.
\end{singlespace}
}
\end{table}

\subsection{Experiments}

A galvanized steel washtub was chosen as a container for the granular medium because it is approximately axi-symmetric (see the first paragraph in this section for the exact dimensions).  
The 11-gallon (41.6 litre) tub size was chosen to be large enough to 
fit 7 evenly spaced accelerometers in a radial line within it and yet be small enough
that it was feasible to fill it with our substrate materials. 
Because the base of the tub has a small protruding lip
on its rim,  we placed it on about a cm deep layer of fine sand that in turn is on a board that lies on the floor of our lab.   The fine sand base reduces motions in the tub's metal base. 
The drop height 
was chosen so that the peak acceleration in the first accelerometer signal neared, 
but did not exceed, the $\pm 3$ g cutoff of the accelerometer integrated circuit.

For each experiment  
we compute the impact velocity $v_{imp} = \sqrt{2 g_\oplus h_{\rm drop}}$   
using the drop height and the projectile kinetic energy 
$K_{imp} = \frac{1}{2} m_p v_{imp}^2 $ from the projectile's mass $m_p$ and impact velocity. 
Here $g_\oplus$ is the gravitational acceleration
on the Earth.
For each experiment we compute dimensionless ratios used for crater
scaling \citep{holsapple93}.
The Froude number depends on the impact velocity, projectile radius $R_p$ and
gravitational acceleration 
\begin{equation}
Fr = \frac{v_{imp}} {\sqrt{g_\oplus R_p}} = \pi_2^{-\frac{1}{2}}, \label{eqn:Fr}
\end{equation}
and is directly related to the dimensionless $\pi_2$ parameter adopted by \citet{holsapple93}.
The ratio of substrate bulk density to projectile density is equal to the dimensionless $\pi_4$ parameter,  
\begin{equation}
\pi_4  = \frac{\rho_s}{\rho_p}. \label{eqn:pi_4}
\end{equation}

Experiments are listed in Table \ref{tab:exps}
with accelerometer coordinate template, 
drop height,  impact velocity and kinetic energy  and the dimensionless parameters $Fr$ and$ \pi_4$. 
The maximum width between the top edges of the crater rim was measured to give the crater 
diameter $D_{cr}$ and this too is listed in Table \ref{tab:exps}.
We compute and list the dimensionless parameter 
\begin{equation}
\pi_R = R_{cr}  \left( \frac{\rho_s}{m_p} \right)^\frac{1}{3}  \label{eqn:pi_R}
\end{equation}
where crater radius $R_{cr} = D_{cr}/2$. 
Repeated experiments with the same projectile,
drop height, substrate and projectile have the same crater diameter and impact velocity. 
This is why in Table \ref{tab:exps} a series of coordinate templates is listed for the 
experiments in millet. 

Experiments are labelled with first letter M if the substrate
is millet and with first letter S if the substrate is sand.
Experiments are also labelled by their coordinate template. For example, ML5 denotes an experiment in millet with accelerometers placed according to the L5 coordinate template.
The SL5 experiment also uses the L5 coordinate template but is into sand. 
Pushing an accelerometer board into sand requires more force than pushing one into millet. It is easier and less damaging to the PCB boards and wiring to embed the accelerometers into millet than sand.  Consequently we do experiments with more coordinate templates in millet than sand. 
We have checked that pulses measured in repeated experiments are similar. 
We estimate that there are variations of about $\pm$ 1/2 cm in radial position and depth
for the accelerometers and variations of about $\pm$ 1/2 cm in the intended site of impact. 

\begin{table}[htbp]  
\caption{Experiments} \label{tab:exps}
\if \ispreprint1
\begin{tabular}{p{24mm} p{16mm} p{18mm}llll}
\else
\begin{tabular}{p{30mm} p{25mm} p{25mm}llll}
\fi
\hline
\multicolumn{2}{l}{Date of  experiments}     & \multicolumn{2}{l}{11/27/2021}    \\
\multicolumn{2}{l}{Container}         &  \multicolumn{2}{l}{41.6 litre tub}    \\
\multicolumn{2}{l}{Floor}               &  \multicolumn{2}{l}{Fine sand }    \\
\hline
Substrate                  &   & Millet           &  Fine sand      \\
Projectile          && Rubber Ball\!& Glass Marble  \\
\multicolumn{2}{l}{Accelerometer coordinates}     & L5,L10,L15 \newline R5,R10,R15         &  L5     \\  
\multicolumn{2}{l}{Experiment labels}   & ML5,ML10\newline ML15,MR5 \newline MR10,MR15\!& SL5 \\
Drop height  & $h_{\rm drop}$(cm)  &  101  & 115   \\
Impact velocity  &  $v_{ imp}$(m/s) & 4.4     & 4.7     \\
Crater diam.  & $D_{cr}\!$ (cm)   & 11 & 7.5    \\
Kinetic energy  & $K_{imp}$ (J) & 0.27  &  0.31  \\
Froude number & $Fr\! =\! \pi_2^{-\frac{1}{2}}$\!\! &  10.5 & 12.8   \\
Inertial ratio  & $\pi_2$ & 0.009 & 0.006 \\
Density ratio  &$\pi_4\! =\!\rho_s/\rho_p$\!\!& 0.70 & 0.57   \\
Crater ratio  &    $\pi_R$          & 1.7 & 2.0  \\
\hline
\end{tabular}
\\ \par {\footnotesize \begin{singlespace}
Notes: Coordinate templates for the accelerometers are listed in Table \ref{tab:template}.
Projectile properties are listed in Table \ref{tab:imp} and substrate properties in Table \ref{tab:granular}.
Experiments are labelled with first letter an `M' if they are into millet and first letter an `S' if they are into sand,
followed by the coordinate template name. 
Dates are written as Month/Day/Year.
Dimensionless numbers $Fr$,  $\pi_2$, $\pi_4$, and $\pi_R$ are defined in 
equations \ref{eqn:Fr}, \ref{eqn:pi_4} and \ref{eqn:pi_R}.
\end{singlespace}
}
\end{table}

\section{Experimental results on pulse peaks}
\label{sec:exp}

\subsection{Accelerometers in a line and at the same depth}
\label{sec:ML5}

In Figure \ref{fig:ML5}
we show signals from an experiment of an impact into millet, denoted experiment ML5, 
and in Figure \ref{fig:SL5} we show a similar experiment into sand, denoted SL5.
The experiments have properties listed in Table \ref{tab:exps}.  
For these two experiments, the 7 accelerometers
were placed at a depth of 5 cm and in a radial line.   
The accelerometer positions 
are described with template L5 with coordinates listed in Table \ref{tab:template}.

In Figure \ref{fig:ML5}, the top two panels show the $\hat {\bf R}$ (cylindrical radial) direction 
acceleration and velocity components and the bottom two panels show the $\hat {\bf z}$ (vertical) direction components.
The acceleration signals have been filtered with a Savinsky-Golay filter
that has a window width of 11 samples which is equivalent to a duration of $1.1 \times 10^{-4}$ seconds.
This duration corresponds to a frequency of 11 kHz which exceeds the upper band-pass limit (1600 Hz)
of the accelerometer outputs, so this filter only removes noise.
The velocity is computed by numerically integrating the acceleration signal.  
We compute a cumulative
sum and multiply by the sampling time (the time between data samples).
From each signal we subtract a constant value so that the initial velocity and
acceleration, just prior to impact, are zero.  In Figure \ref{fig:ML5} each accelerometer is plotted
with a vertical offset so that the signals are equally spaced and offset in order of distance from the impact site. Each accelerometer is 
plotted with a different color line. The topmost signal, plotted in red, is the one 
nearest the impact site.  

In Figure \ref{fig:ML5},
the horizontal time axes have been shifted so that zero corresponds to our best estimate
of the impact time.   
We estimated the time of impact by comparing high speed video
with the accelerometer signals, and with both camera and oscilloscopes triggered by the IR break-beam sensor.  We found that the signal starts to rise in the nearest accelerometer  
1 to 2 ms after the projectile first touches the substrate surface.
The impact time in each experiment, relative to one another, was estimated 
from the moment the acceleration
starts to rise in the accelerometer nearest impact 
and by comparing the signals from other experiments done 
with accelerometers at similar locations.    
Unfortunately the impact site varies
slightly from impact to impact.  We suspect variations in the solenoid/projectile contact position and resulting variation in the time 
the IR-break-beam sensor is blocked are responsible for 
an uncertainty of about 1 ms in our estimate for the time of impact.  As the data from all 7 accelerometers was taken simultaneously, the relative timing error between accelerometers from the same
experiment is equal to the sampling time or $10^{-2}$ ms.

As expected, the pulses shown in Figures \ref{fig:ML5} and \ref{fig:SL5} 
are strongest for the accelerometer nearest the impact site in both radial and vertical acceleration   
velocity components.  
Comparison of the first three accelerometer profiles (those nearest the impact site) show that 
the pulses drop in amplitude, and travel 
away from the impact site in the radial direction.   
The shape of the radial component of acceleration resembles those
seen in other impact experiments into granular media \citep{mcgarr69,yasui15,Matsue_2020}.

In the accelerometer nearest the impact site plotted in red in Figure \ref{fig:ML5},  
the propagated pulse begins with a negative $z$ acceleration component.  
This accelerometer has a depth of 5 cm and a similar radius $R$ in cylindrical coordinates.
If the seismic source caused by the impact lies near the surface then we
expect a longitudinal or pressure wave that is traveling
radially outward from the impact site (in spherical coordinates).  This would give a negative
$z$-component of acceleration.   
Because the downward pulse in the first signal is due to the depth of the accelerometer,
the $z$ acceleration components should not be interpreted in terms of a Rayleigh wave.
We discuss the directions of the motions in more detail in Section \ref{sec:ray}.

A comparison between 
Figure \ref{fig:ML5} and Figure \ref{fig:SL5} shows that pulses have a similar shape in sand and millet,
though the pulses are narrower in the sand.   We will discuss pulse duration in more detail in Section \ref{sec:duration} below.

\subsection{Accelerometers at the same radius and at different depths}
\label{sec:MR5}

\begin{figure*}[htpb] \centering 
\includegraphics[width = 6.2 truein,trim = 0 0 0 0, clip]{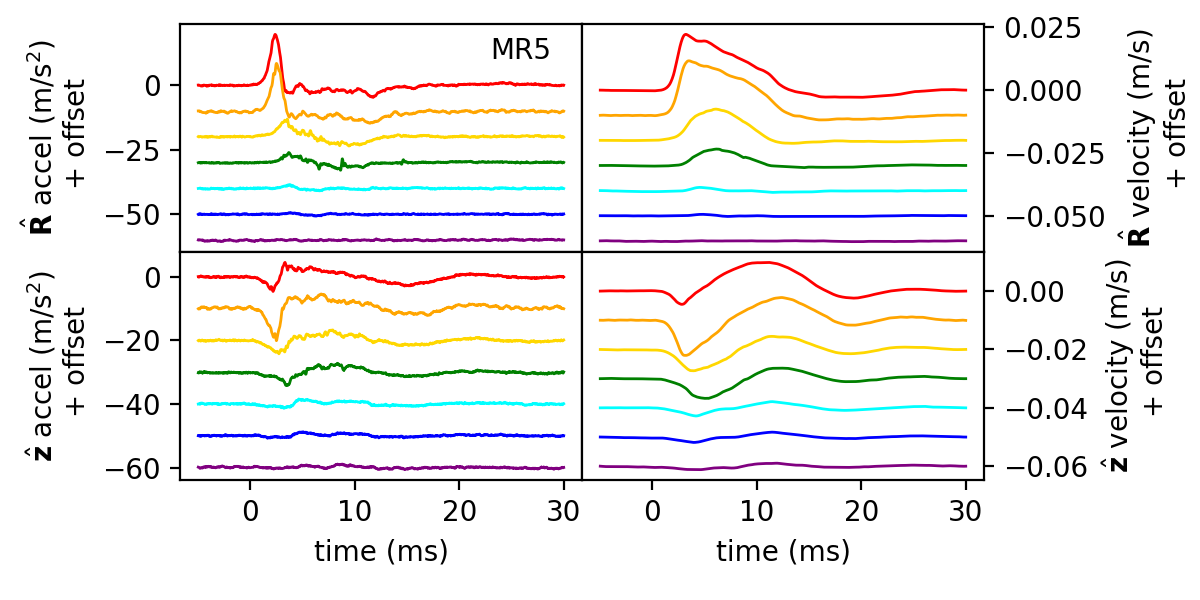}
\caption{Similar to Figure \ref{fig:ML5} except we show the MR5 experiment data, in  millet, 
where the accelerometers
are arranged at the same cylindrical radius $R$ but at different depths.   
 \label{fig:MR5}
}
\end{figure*}

Since we have difficulty embedding the accelerometers at larger depths
in the sand (as discussed at the end of Section \ref{sec:exp}), we use experiments in millet to explore the depth dependence of pulse propagation.
In Figure \ref{fig:MR5} we show the MR5 experiment, into millet, where all 
accelerometers are at the same cylindrical radius $R$ but at different depths.
A comparison between Figure \ref{fig:ML5} (also into millet but with all
accelerometers at the same depth) and Figure \ref{fig:MR5} suggests
that pulse strength decays less rapidly with depth than with cylindrical radius $R$.
The shallowest accelerometer that is nearest the impact site, shown in red in Figure \ref{fig:MR5},
has a pulse that is similar in strength to the second one that is deeper and
shown in orange.  The comparison between Figure \ref{fig:ML5} and Figure \ref{fig:MR5}
implies that propagation is not spherically symmetric about the impact site. More energy propagates downward than horizontally outward.  
Similar phenomena was seen in two-dimensional simulations of oblique impacts into granular media \citep{Miklavcic_2022}. 
In these simulations, 
plastic deformation extends further laterally than vertically, causing a more rapid decay of energy in the lateral pulse compared to the vertical pulse.
In Section \ref{sec:pk} and using 2d maps of peak velocity 
and peak acceleration we see more evidence for angular dependent pulse amplitudes. 
Pulse peak accelerations and velocities are discussed and shown in more detail 
as a function of distance from the site of impact in section \ref{sec:amp}. 

\begin{figure}[htbp] \centering 
\if \ispreprint1
\includegraphics[width = 3.3 truein, trim = 0 10 0 0, clip]{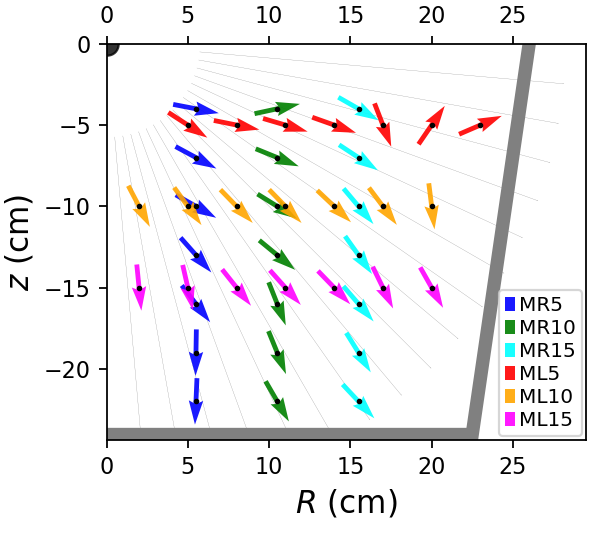}
\else
\includegraphics[width = 4.3 truein, trim = 0 10 0 0, clip]{nov27theta_a.png}
\fi
\caption{Ray angles are shown as colored arrows.  The angles are
measured from the ratio of vertical to horizontal 
acceleration components at the moment of peak acceleration 
for the ML5, ML10, ML15, MR5, MR10 and MR15 
experiments in millet.  Each experiment is shown with a different color arrow.
Accelerometer locations are shown with black dots.  Plot axes are in cylindrical coordinates, with the site of impact at the origin which is marked with a black quarter circle. 
The outline of the tub is shown with a thick grey line.  Thin grey lines show radial rays originating from the site of impact. 
 \label{fig:theta}} 
\end{figure}

\subsection{Ray angles}
\label{sec:ray}

Using the accelerometer signals and coordinates of each accelerometer, we compute 
the radial components of acceleration $a_r = {\bf a} \cdot \hat {\bf r} $  and velocity  $v_r = 
{\bf a} \cdot \hat {\bf r} $ where unit vector 
$\hat {\bf r}  = ( R \hat {\bf R} + z \hat {\bf z})/\sqrt{R^2+ z^2}$.   From the radial components of acceleration,  we measure 
the time of peak acceleration, which we denote $t_{a,prop}$ as it represents a travel or propagation time, and the value of the radial component of the acceleration at this time, $a_{r,pk}$. 
At the time of peak radial velocity $t_{v,prop}$ we measure the value of the peak radial velocity component $v_{r,pk}$.  

In Figure \ref{fig:theta} we show the direction of
the acceleration vector $\frac{\bf a}{a}$ at the locations of each accelerometer
in the millet experiments computed at the time $t_{a,prop}$.  The black dots show the locations
of the accelerometers and at each location a unit vector is plotted
that shows the acceleration direction.  The origin is the site of impact. 
Arrows in Figure \ref{fig:theta} are nearly radial from the impact point.  Using all the points in Figure \ref{fig:theta}, we compute the 
 standard deviation of the acceleration angle 
subtracted by the angle of a ray originating from the origin and find that it is $22^\circ$.  
Discarding the 7-th and most distant accelerometer positions in each experiment, 
the standard deviation is $20^\circ$.
Variations in accelerometer orientation and position and in the location of the impact can account for most of this scatter. 
Because the arrows in Figure \ref{fig:theta} are nearly radial from the impact point,  
we infer that pulse propagation is primarily longitudinal.  
Large deviations are seen at large radius $R$ where the signals are weak and where reflections
from the tub wall at later times could have influenced the direction of 
wave propagation.  
Since propagation rays are nearly radial, the wave propagation velocity
cannot be strongly dependent upon depth. 
Angles computed at the time of peak velocity $t_{v,prop}$
from the velocity components 
are similar to those computed at peak acceleration using the acceleration components.

\subsection{Peak accelerations and velocities}
\label{sec:pk}

\begin{figure}[htbp] \centering 
\includegraphics[width = 3.2 truein, trim = 25 5 30 5,  clip]{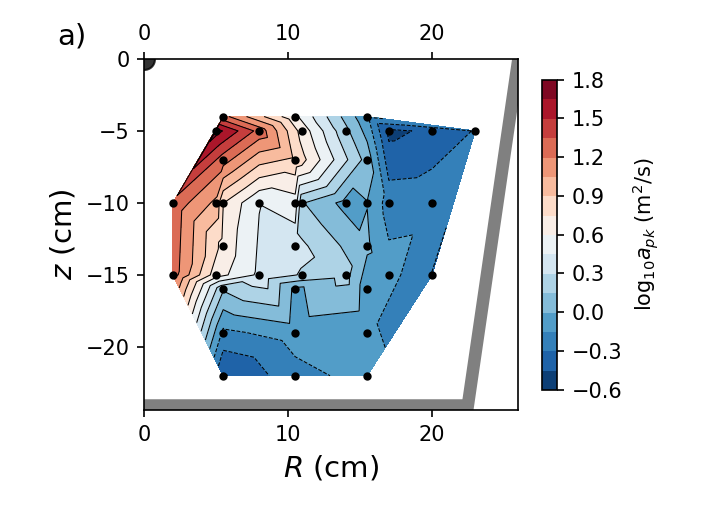}
\includegraphics[width = 3.2 truein, trim = 25 5 30 5, clip]{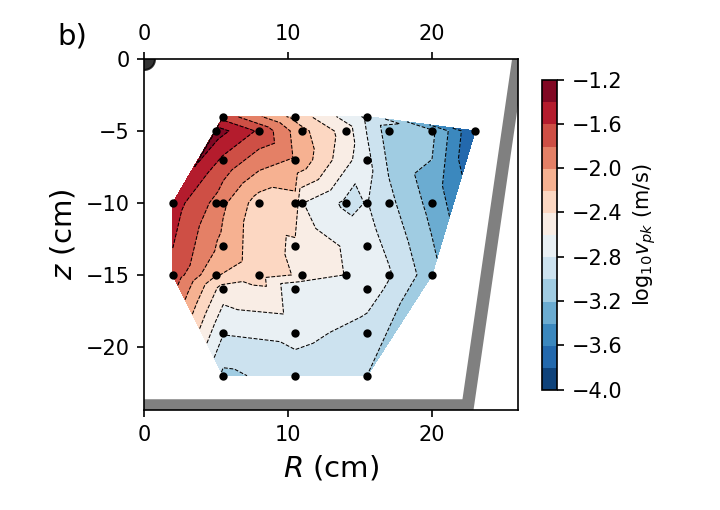}
\caption{a) Peak pulse radial acceleration as a function of 
accelerometer position for the millet experiments. The thick grey line shows the outline of the tub. The black quarter circle shows the site of impact. 
b) Similar to a) except we plot peak radial velocity components. 
\label{fig:pk_2d}} 
\end{figure}

Because the acceleration directions imply that the pulses in our experiments are longitudinal pressure waves, we compute additional 
 quantities using the radial ($r$) components of acceleration and velocity.
In Figure \ref{fig:pk_2d},  
peak radial components of acceleration and velocity are plotted with contour plots
 as a function of accelerometer position. 
In this figure, the locations of the accelerometers
are shown with black dots.  
The contour plots were made in python with a triangulation 
routine (tricontour) for irregularly sampled data.
Figure \ref{fig:pk_2d} shows that pulse propagation is not spherically symmetric about the impact point.  
We estimate that peak amplitudes are about twice as large directly below
the impact as at the same distance from impact but along a direction of 45$^\circ$ from vertical.

\begin{figure}[htbp] \centering 
\includegraphics[width = 3.3 truein,trim =20 10 0 0, clip]{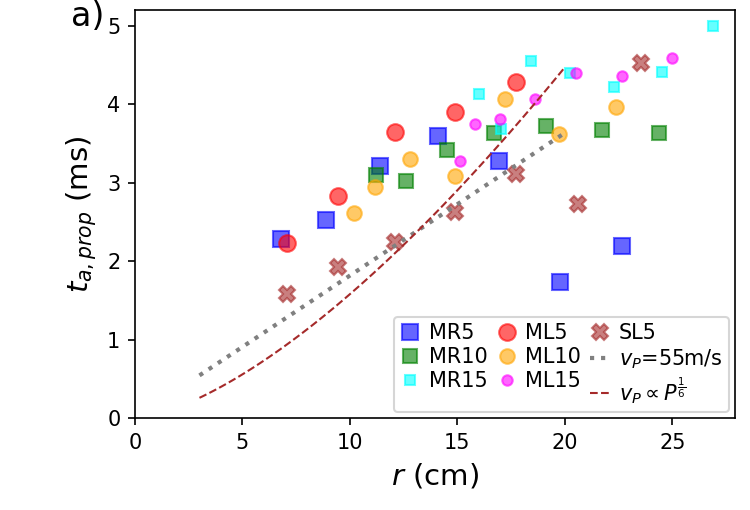}
\includegraphics[width = 3.3 truein,trim =20 10 0 0, clip]{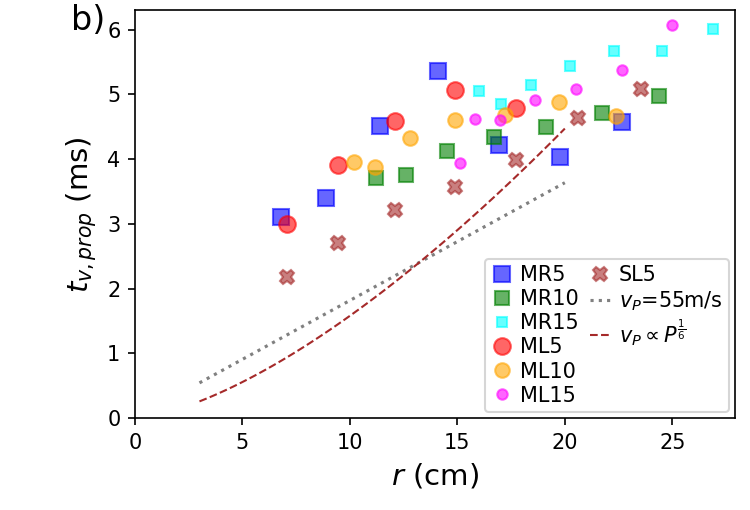}
\caption{a) 
Time of peak radial acceleration $t_{a,prop}$, measured from the time of impact versus distance between impact site and accelerometer location ($r$). 
Each experiment comprised of 7 accelerometers is plotted with a different color marker and a point is plotted for each accelerometer in that experiment. 
Experiments into millet that have accelerometer positions at a single cylindrical radius $R$ are shown with square markers.   
Experiments into millet that have accelerometer positions at a single depth 
 are shown with round markers.  
The SL5 experiment into sand is shown with brown X-shaped markers.
The dotted grey line has a slope giving pulse travel speed of 55 m/s.
The dashed brown line shows a model with pulse velocity sensitive to pulse amplitude. 
b) Similar to a) except we plot time $t_{v,prop}$ of peak radial velocity, $v_r$, as a function of distance
from impact site.  
\label{fig:speeds}}
\end{figure}

\subsection{Pulse travel speed}
\label{sec:speed}

For the SL5 experiment into sand and the six experiments in millet (ML5, ML10, ML15, MR5, MR10, and MR15)
we mark the times of the peak radial acceleration $t_{a,prop}$ in each accelerometer and plot them versus distance $r$ from the point of impact.   
Each experiment is plotted in Figure \ref{fig:speeds}a with points that have unique color and shape.  Figure \ref{fig:speeds}b is similar except it shows
the time of peak radial velocities,  $t_{v,prop}$.

In Figure \ref{fig:speeds}a and b 
we have also plotted a grey dotted line that corresponds to a travel velocity of $v_P = 55 $ m/s.
On these plots a steeper line corresponds to a slower travel speed. 
The 55 m/s speed is approximately consistent with both radially and depth distributed
accelerometers in millet and with the radially distributed accelerometers in fine sand.
The travel speed below the surface (as measured from the MR5 experiment)
is similar to that near the surface (as measured in the ML5 experiment).
Surprisingly the pulse travel speed in fine sand is similar to that in millet.

A line going through the points showing millet experiments in Figure \ref{fig:speeds}a,  
 does not pass through the origin, rather
it would intersect $r=0$ at about 1 ms.
In Figure \ref{fig:speeds}a, we plot the time of peak acceleration, not the time when
the acceleration first begins to rise. 
The time $t=0$ is approximately 
the time when the projectile first touches the substrate surface. 
Examination of high speed video shows that 
the pulse is launched quite early, less than 2 ms after the projectile first touches the surface.
In high speed video, motions on the surface can be seen immediately after impact with a front that moves rapidly away
from the impact site before most of the ejecta curtain is launched and obscures the surface. 
Falling at a speed of about 5 m/s, it takes the
projectile about 2 ms to drop 1 cm, a distance approximately equal to the projectile radius.
The times shown in Figure \ref{fig:speeds} are those of pulse peaks, so a 1 or 2 ms delay would be 
consistent with the pulse peak arriving somewhat later than the rising pressure wave that is launched when the projectile first contacts the substrate surface.

Hertzian contact models predict a power-law dependence of 
the effective pulse travel speed $v_P$
on ambient or confinement pressure $P_0$ \citep{Duffy_1957,Liu_1992,Johnson_2000,Somfai_2005} 
with a scaling of $v_P \propto P_0^{\beta_0} $  with index $\beta_0\approx  \frac{1}{6} $. 
Predictions for the index vary from 1/4 to 1/6 for models of 
one-dimensional (1D) chains \citep{Herbold_2009} 
and three-dimensional (3D) ordered \citep{Gilles_2003,Coste_2008} sphere packing.
Experiments measuring a similar range in the index  \citep{Tell_2020,Zhai_2020}.
Note that $v_P \ll c_g$  where $c_g$ is the sound speed in the grain's material. 
The travel speed of a pressure pulse may also depend on the pulse amplitude or peak pressure $P_{pk}$ with a power law scaling,   
$v_P \propto P_{pk}^{\beta_P}$ with index $\beta_P \approx 1/6$ and this is predicted
theoretically and observed in experiments   \citep{Gilles_2003,vandenWildenberg_2013,Santibanez_2016,Tell_2020}.
As the pulse pressure amplitude decreases, $P_{pk} \lesssim P_0$,  the pulse propagation speed 
undergoes a transition from a nonlinear and shock-like propagation regime, where 
the speed depends on the peak pressure, to a linear propagation regime  
where the propagation speed depends on the ambient or confinement pressure \citep{vandenWildenberg_2013,Santibanez_2016,Tell_2020}.
To show the possible dependence of travel speed on
pressure,  
in Figures \ref{fig:speeds}a,b we also show a dashed brown line
which has pulse speed dependent on pressure in the pulse to the 1/6 power.
Pulse peak pressure and pulse travel speed are estimated using equations  \ref{eqn:1_6}
and \ref{eqn:P_peak2} and using the model described in more detail in section \ref{sec:DART}.

There are deviations at larger distances from the impact point with 
pulses seen at the most distant accelerometers having shorter pulse travel times compared
to those predicted by a constant speed homogeneous medium.
There are also deviations at shorter distances  $r<15 $ cm from the impact site 
with longer travel times than estimated with a constant speed model.

We consider possible explanations for deviations from a linear relation between travel
time and distance to the impact site. 

1) Pulse travel speed could be faster with increasing depth. This would be expected
if pulse travel speed is set by the strength of contacts and the contact
forces depend upon hydrostatic pressure \citep{Duffy_1957,Walton_1987,Liu_1992,Johnson_2000}.  
In this case we would measure
a shorter travel time for distant accelerometers, compared to that predicted with a homogeneous 
model.  Wave propagation rays would be curved. 
The more distant accelerometers in the ML5 and SL5 experiments would
see pulses arriving from below, giving positive $z$ velocity 
and acceleration components.

2) The pulse travel speed depends on the pressure in the pulse itself 
(e.g., \citealt{vandenWildenberg_2013}).
In this case the travel speed decreases as a function of travel distance because
the pulse amplitude decays as it travels.  We would measure
a longer travel time for more distant accelerometers, compared to that predicted with a homogeneous model.  Such a model is illustrated with the brown dashed lines in Figure \ref{fig:speeds}. 


3) Because the pulses are broad, reflections can affect the measurement of the peak time.
This primarily affects the accelerometers nearest the tub edge or bottom.
If a pulse is reflected from a hard edge, a positive pressure pulse is reflected 
as a positive pressure pulse, but the direction of travel reverses, giving the opposite
sign in acceleration and velocity and truncating the later part of the pulse.  
The estimated peak time of a broad pulse might be reduced
by the reflected wave.
 
Because the time of travel decreases at larger distances, rather than increases, 
we infer that the mean travel speed could be faster for longer distances traveled.
This is consistent with possibility 1, where the travel speed depends on depth
and is faster below the surface. 
However, if the speeds are faster at depth, propagation rays would approach
the accelerometers from below, giving a z-component in the accelerations for
the most distant accelerometers and this is not seen in the MR5 or SR5 data sets and is ruled out by 
 the acceleration directions shown in Figure \ref{fig:theta}.  We discard possibility 1.


The peak pressure
dependent travel speed (possibility 2) gives increasing mean travel times as a function of
distance compared to a homogeneous model,  
where travel time is proportional to travel distance.  This could be consistent with the arrival times
for the accelerometers that are about 12 cm of the impact site 
which seem high for the MR5 and ML5 experiments, 
but would not be consistent with the relatively short travel times for the most distant accelerometers. 
Possibility 2 (pressure dependent pulse propagation velocity) could account for the relatively 
higher travel times at $r \sim 15$ cm  but cannot account for
the relatively shorter travel times at $r>15$ cm from impact site.

With a travel speed of 55 m/s it takes a wave only about 5 ms to travel from impact site 
on the surface horizontally to the tub edge
or from impact site vertically to the tub base.  Because the amplitudes
drop rapidly as a function of distance from impact site, reflections would primarily affect 
pulse peaks seen in the accelerometers most distant from impact site.
The deviations from radial acceleration directions at locations most distant from the
impact site seen in Figure \ref{fig:theta} also support this interpretation.  
Reflections off the tub walls (possibility 3) are the most likely explanation for the flattening of
the estimated peak arrival times in the most distant accelerometers. 

In summary, our pulse peak arrival times would be consistent with a 
pulse travel speed that is somewhat higher near the impact site than a constant velocity model
due to a pressure dependence in the pulse propagation velocity.   We test this possibility further
by estimating the pressure in the pulses to see whether they exceed hydrostatic pressure 
in Section \ref{sec:pressure}.
We suspect that
reflections have affected our peak time measurements for the most distant accelerometers which
have the weakest and noisiest signals. 


An estimate for the pulse travel speed is useful to estimate physical 
quantities such as the pressure amplitude of the pulse and the seismic energy efficiency.
We estimate a travel speed of $v_P \sim 55$ m/s for both sand and millet, based
on our estimate of pulse peak arrival times discussed here and we use this value in our discussions below.

\begin{table}[htbp] 
\caption{Peak values and related quantities} \label{tab:pressure}
\if \ispreprint1
\begin{tabular}{p{32mm}p{14mm}p{8mm}p{8mm}p{9mm}llll}
\else
\begin{tabular}{p{42mm}p{18mm}p{12mm}p{12mm}p{14mm}llll}
\fi
\hline
                                 &                           &  MR5    & ML5 & SL5 \\
 \hline
Peak radial accel.  &$a_{pk}$(m$^2\!$/s) &  17.0 & 51.9 & 43.0 \\
Peak radial velocity   &$v_{pk}$(m/s)           &  0.020 & 0.051 & 0.026 \\
\hline
Adopted pulse speed  & $v_P$ (m/s) & \multicolumn{2}{c}{55}   & 55 \\
Speed ratio                  &  $v_P/c_g$  &   \multicolumn{2}{c}{$\sim\!$0.2}&$\sim\!$0.03\\
\hline
Pressure perturbation\!\!\!& $P_{pk}$(Pa)  &  840 & 2100 & 2200  \\
Hydrostatic depth        & $H_p$(cm)  & 11 & 28 & 15 \\
Distance to impact\!\!\!& $r$ (cm) &   6.8 & 7.1 & 7.4 \\
Distance ratio            &  $r/R_{cr}$ & 1.2 & 1.3 & 1.9 \\
Pressure ratio           & $P_{pk}/E_g$  &8$\times\!10^{{-}6}$&2$\times\!10^{{-}5}$&2$\times\!10^{{-}7}$\\
Seismic Energy   & $E_{seis}$(mJ) &  2 & 9 & 5 \\
Seismic efficiency & $k_{seis}$(\%) & 0.8 & 3.3 & 1.5 \\
\hline
\end{tabular}
\\
\par { \footnotesize \begin{singlespace}
Notes:  We list the peak radial velocity and 
 peak radial acceleration in the accelerometer
nearest the impact site in the MR5 and ML5 experiments into millet
and in the SL5 experiment into sand.  The radial distance from impact site $r$. 
The pressure perturbations are
estimated using Equation \ref{eqn:P_peak} using the peak radial velocity component. 
Sound travel speed $c_g$ and elastic modulus $E_g$ within a grain are
taken from Table \ref{tab:granular} and used to compute the speed ratio $v_P/c_g$
and the pressure ratio $P_{pk}/E_g$.
The depths $H_p$ are where peak pulse pressure would equal hydrostatic pressure
and are estimated with Equation \ref{eqn:H_p}.
The distance ratio $r/R_{cr}$ is that of the nearest accelerometer from the impact point
divided by the crater radius (listed in Table \ref{tab:exps}).
The seismic energy $E_{seis}$ is computed with Equation \ref{eqn:E_seismic} and the
seismic efficiency is computed with  
$k_{seis} = E_{seis}/K_{imp}$ and kinetic energies listed
in Table \ref{tab:exps}.  Both quantities are computed for the accelerometer nearest
the impact site. 
\end{singlespace}
}
\end{table}

\subsection{Estimates for the peak pulse pressure} 
\label{sec:pressure}

The pressure and velocity perturbations  in a sound wave are related via 
\begin{equation}
dp \sim \rho c_s dv , \label{eqn:dp}
\end{equation} 
where $c_s$ is the sound speed and $\rho$ is the density of the medium.
\citet{vandenWildenberg_2013} found this relation is also obeyed for pulses propagating in a granular medium 
but after replacing the sound speed with an effective sound speed for pulse propagation $v_P$ and using 
density $\rho_s$,  the bulk density of the granular medium.
With the peak pressure in a pulse $P_{pk}$,  
and $v_{pk}$ the peak velocity, 
Equation \ref{eqn:dp} becomes 
\begin{align}
P_{pk} &\sim \rho_s v_P v_{pk}. \label{eqn:P_peak} 
\end{align}

Using Equation \ref{eqn:P_peak}, we estimate the size of the pressure peaks in our pulses from 
the peak velocities seen in the integrated acceleration signals.
We estimate the peak pressure using the velocity peak seen in the accelerometer
nearest the impact site for three experiments,  MR5, ML5 and SL5 
with the matching substrate density for millet or sand (listed in Table \ref{tab:granular}) 
and our estimate for the pulse travel speed (listed in Table \ref{tab:pressure}).
The estimated peak pressures are about 1 kPa and are listed in Table \ref{tab:pressure}. 

Since hydrostatic pressure depends upon depth $H$, with $P_0(H)  = \rho_s g H$ we ask, at what depth $H_p$ does the peak
pressure match the hydrostatic pressure?  
The depth where peak pressure matches hydrostatic pressure, $P_0(H_p) = P_{pk}$,
is  
\begin{equation} 
H_p = \frac{P_{pk}}{\rho_s g_\oplus}.  \label{eqn:H_p}
\end{equation}
These depths  
are also listed in Table \ref{tab:pressure} using
the peak pressures measured in the accelerometers
nearest impact for MR5, ML5 and SL5 experiments.  
The depths are 28 and 15 cm for the ML5 and SL5 experiments. 
Because the pulse height rapidly decreases as the pulse travels, 
our estimated pulse pressures exceed the hydrostatic
pressure,  $P_{pk} > P_0$ only near the impact site and near the surface.
However, the distance at which the transition occurs is similar to $H_p$.  
As the depth $H_p$ lies within our substrate container, the transition between pulse pressure dominated and hydrostatic pressure
 dominated could account for variations
in travel time we discussed in Section \ref{sec:speed}.

The ambient and hydrostatic pressure on a low-g environment such as an asteroid is low, so pulses arising from all but
the lowest energy impacts would be in the non-linear pulse propagation regime \citep{Sanchez_2021}. 
Near (within about 10 cm of) the surface and the impact site, we estimate that our experiments are just barely in a regime that might apply to low-g environments.

\subsection{The pulse travel speed}
\label{sec:speed_value}

We were surprised to measure similar pulse travel speeds in sand and millet.  In this subsection we compare these speeds to those measured in other experiments.

\cite{yasui15} measured a pulse travel speed of 109 m/s in 200 $\mu$m diameter glass
beads, whereas \cite{Matsue_2020}  measured a pulse travel speed of 53 m/s into
quartz sand and similar to ours.  

The wave front propagation speed within granular media
is dependent on front pressure, confining or ambient pressure, and whether the substrate is organized in a lattice or is disordered \citep{Somfai_2005,vandenWildenberg_2013}.  
A polydisperse and disordered granular medium can effectively have a lower bulk modulus and so 
a lower pulse propagation speed
compared to a similar but monodisperse medium since smaller grains do not tend to carry strong 
forces within the force chain network \citep{Petit_2018}.

To compare our pulse propagation speeds to models and other experiments  (as shown in 
Figure 8 by \citealt{Somfai_2005}),  we compute the speed ratio $v_P/c_g$ which is the pulse travel speed $v_P$ divided by that estimated for the grain material $c_g$,  and we compute a pressure ratio $P_{pk}/E_g$ which is the peak pulse pressure $P_{pk}$ divided by the elastic modulus of the grains $E_g$. 
We use peak pressures as they are similar in size to the hydrostatic pressure in the middle of the tub.  
We have checked that the relation for $v_P/c_g$ vs $P_0/E_g$, using confinement pressure, measured in disordered glass spheres  by \citet{Jia_1999} is consistent with (to within an order of magnitude) the relation between $v_P/c_g$ and $P_{pk}/E_g$, using peak pressure, found by \citet{vandenWildenberg_2013}. 
Using the values for the sound speed within the grains (listed in Table \ref{tab:granular}), we estimate $v_P/c_g \sim 0.15$ in millet and 0.03 in the fine sand. 
Using the elastic moduli listed in Table \ref{tab:granular} we compute 
$P_{pk}/E_g$ from the peak pressures estimated in the MR5, ML5 and SL5 experiments
and list the resulting values, along with ratio $v_P/c_g$ in Table \ref{tab:pressure}.  
The pressure ratio $P_{pk}/E_g  \sim 10^{-5}$ in millet and $\sim 10^{-7}$ in sand.

Though we were surprised that the pulse propagation speeds are similar in sand and millet, 
on a plot of $v_P/c_g$ vs $P_{pk}/E_g$, our measurements are approximately
consistent with experimental measurements in disordered glass spheres by \citet{Jia_1999}.
Monodisperse grains in lattice structures tend to have higher propagation
speeds, (as shown in Figure 8 by \citealt{Somfai_2005}) suggesting that our low pulse propagation speeds are consistent with those predicted and measured in disordered granular media.

\subsection{Seismic energy flux}
\label{sec:seismic}

\begin{figure}[htbp] \centering 
\if \ispreprint1
\includegraphics[width = 3.5 truein, trim = 15 5 20 5, clip]{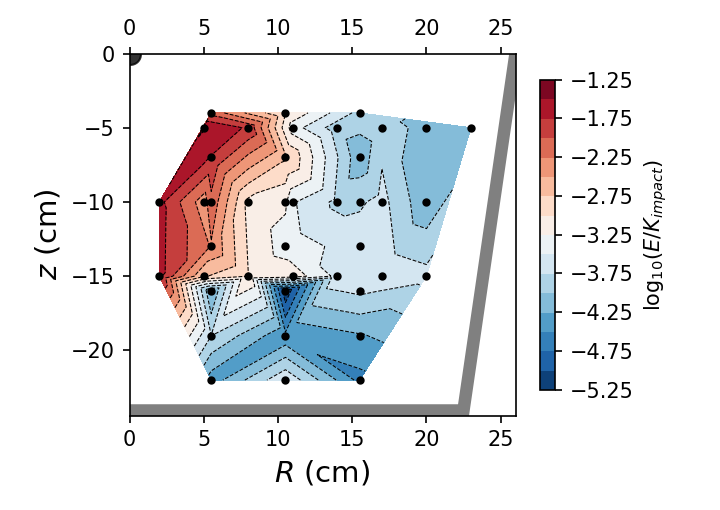}
\else
\includegraphics[width = 4.5 truein, trim = 15 5 20 5, clip]{nov27EJ_2d.png}
\fi
\caption{Similar to Figure \ref{fig:pk_2d} except
we show the pulse seismic energy as a function of accelerometer location.  
Seismic energy is 
computed using Equation \ref{eqn:E_seismic} using
the velocity signals from each accelerometer in the six millet experiments. 
More energy propagates downward than horizontally. 
\label{fig:EJ_2d}
}
\end{figure}

We estimate the energy in the seismic pulse by integrating the radial velocity signal 
in an accelerometer  $v_{r}(t)$ 
\begin{equation}
E_{seis}(r) \approx \int_0^{t_1} dt\ \rho_s v_{r}(t)^2  2 \pi r^2  v_P  \label{eqn:E_seismic}
\end{equation}
following \citet{yasui15}.  The radius $r$ is the distance between the
impact site and the accelerometer.  In Equation \ref{eqn:E_seismic}
we have assumed that the energy flux at the radius of
the accelerometer is approximately $\rho_s v_r(t)^2 v_P$ and 
equipartition between elastic and kinetic energy (supported by \citealt{vandenWildenberg_2013}).
In Equation \ref{eqn:E_seismic}, 
we ignore the sensitivity of pulse amplitude to angle from vertical.  
We integrate from $t=0$ (time of impact) to $t_1 = 10$ ms so that some  energy in reflections is excluded,
yet we capture most of the pulse that travels from the impact site. 

The seismic efficiency $k_{seis}$ is estimated 
with the kinetic energy of the impact, 
$k_{seis} \equiv E_{seis}/K_{imp}$.
Seismic energies and efficiencies are computed using Equation \ref{eqn:E_seismic} for the MR5, ML5 and SL5 experiments for the 
accelerometers nearest the impact site and 
are listed in Table \ref{tab:pressure}.  These quantities are computed using our estimate for the pulse travel speed $v_P$, also
listed in that table, the substrate densities listed in Table \ref{tab:granular} and 
projectile kinetic energies that are in Table \ref{tab:exps}.

If attenuation is rapid, then an estimate 
for the seismic efficiency would be sensitive to distance from impact point.
Our seismic efficiencies are about a percent and are larger than those computed by \citet{yasui15} 
who found $k_{seis} \sim 10^{-4}$ at a distance of $r/R_{cr} = 4$
and $k_{seis} \sim 5 \times 10^{-4}$  at a distance of $r/R_{cr} = 1$.
Here $r/R_{cr}$ is the ratio of the distance between the accelerometer and impact site
and the crater radius.  In comparison, 
our seismic efficiencies are computed for accelerometers with distance from impact 
$r/R_{cr} \sim 1.2$ and 2.  
As our impact velocities are lower than those of \citet{yasui15}, 
our surprisingly large seismic efficiencies support the suggestion by \citet{yasui15} that seismic efficiency is sensitive to the energy of impact.  
At high impact energy
energy is lost via shock heating and fracture.  
At low impact energy 
the fraction of energy lost could be dependent on the impact energy if attenuation is sensitive to seismic pulse pressure \citep{vandenWildenberg_2013}.

In Figure \ref{fig:EJ_2d} we plot seismic energy estimates as a function of position
within the medium.   For each accelerometer in the millet experiments
we use Equation \ref{eqn:E_seismic} to estimate the total seismic energy at the radius
$r$ of the accelerometer.  Equation \ref{eqn:E_seismic} does not take into account
the angular dependence of pulse propagation so seismic energies estimated below the surface
are higher than those nearer the surface.  The angular dependence resembles that of the peak velocity map shown in Figure \ref{fig:pk_2d}b.

\begin{table}[htbp] 
\caption{Durations} \label{tab:duration}
\if \ispreprint 1
\begin{tabular}{p{32mm}p{16mm}p{17mm}p{6mm}lllll}
\else
\begin{tabular}{p{42mm}p{18mm}p{19mm}p{9mm}lllll}
\fi
\hline
 \ \ \     &         &  MR5, ML5 & SL5 \\
\hline
Pulse duration acc. & $\Delta t_{a} $ (ms) & 1.09, 0.91 & 0.58 \\
Pulse duration vel.     & $\Delta t_{v} $ (ms) & 4.23, 3.46 & 2.26 \\
Normalized dur. acc. & $W_a$  & 0.52, 0.41 & 0.37\\
Normalized dur. vel. & $W_v$  & 1.45, 1.15 & 1.03\\
\hline
Seismic source dur. & $t_{Rss}$ (ms) & 1.0  &0.7 \\
\hline
\end{tabular}\\ { \footnotesize \begin{singlespace}
\par Notes:
We list pulse durations $\Delta t_a, \Delta t_v$ 
from the accelerometers nearest the impact site
from the MR5 and ML5 experiments into millet and the SL5 experiment into sand.
Normalized pulse durations $W_a, W_v$ are the ratio of pulse width 
 to travel time.
The normalized pulse width $W_a$ is computed using the positive portion of the acceleration
pulse and $W_v$ is computed using the positive portion of the velocity pulse.
The crater seismic source duration $t_{Rss}$ is computed using Equation \ref{eqn:tRss}, the crater
radii listed in Table \ref{tab:exps} and the pulse travel speed $v_P$ listed in Table \ref{tab:pressure}.
\end{singlespace} 
}
\end{table}

\subsection{Pulse durations}
\label{sec:duration}

We estimate pulse duration $\Delta t_a$ by measuring the FWHM (full width half max) of the first positive region in the radial component of acceleration $a_r$ which 
peaks at time $t_{a,prop}$.   
Similarly we measure pulse duration $\Delta t_v$ by measuring  
 the FWHM of the first positive region in the radial component of velocity, $v_r$, which 
peaks at time $t_{v,prop}$.   The pulse durations for the accelerometers nearest
the impact site are listed in Table \ref{tab:duration} using the MR5, ML5 experiments into millet and SL5 experiment into sand.  

\begin{figure}[ht]\centering 
\includegraphics[width = 3 truein, trim = 0 0 0 0,clip]{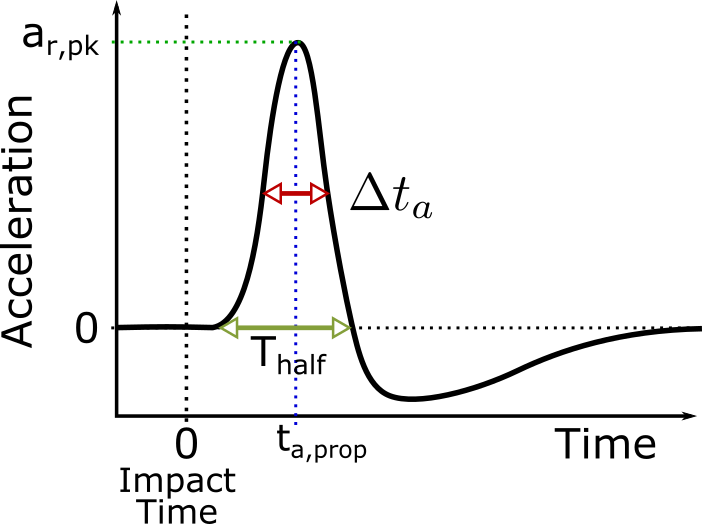}
\caption{A schematic illustration of an impact induced pulse.  We show the measured time $t_{a,prop}$ of peak acceleration $a_{r,pk}$,
and the pulse duration $\Delta t_a$.  We also show $T_{\rm half}$ used
by \citet{yasui15} and \citet{Matsue_2020} to characterize pulse duration. \label{fig:times_illust}}
\end{figure}

A schematic of a impact induced pulse is shown in Figure \ref{fig:times_illust} along
with our measurements for the pulse duration $\Delta t_a$, time of peak $t_{a,prop}$, and peak
acceleration $a_{r,pk}$.
Our peak time $t_{a,prop}$  and our peak acceleration $a_{r,pk}$ correspond to 
$t_{max}$ and $g_{max}$, respectively,  measured by \citet{yasui15} (see their  Figure 3b) and \citet{Matsue_2020} (see their Figure 6e), also in granular substrates but with higher velocity projectiles.   
Our measured pulse duration $\Delta t_a$ is about
half of the duration of the positive peak, $T_{\rm half}$, measured by \citet{yasui15}
and \citet{Matsue_2020}.  

The positive radial acceleration pulse seen in the accelerometer nearest
the impact site shown in Figure \ref{fig:ML5} has FWHM of about 
$\Delta t_a \sim 1$ ms in the ML5
experiment into millet.  However, the FWHM in the similar SL5 experiment into sand
has a shorter FWHM of about half the size. 
The duration of our acceleration pulses is similar to the parameter $T_{\rm half} \sim 0.72 \pm 0.20$ ms measured by \citet{yasui15} for $\sim$ 100 m/s impacts into glass beads that are similar in size to our sand grains. Our pulse durations
are also similar to those for 0.2 to 7 km/s velocity impacts into quartz sand measured by  \citet{Matsue_2020}. 

\citet{yasui15} considered three processes to account
for the duration of a pressure pulse excited by an impact into a granular medium. 

1)  The time for a pressure wave moving at the sound speed to traverse the projectile twice $t_{D,prop}$.
This time is analogous to the time for a shock to propagate forward and a rarefaction wave to propagate backward across the projectile following a high velocity impact \citep{melosh89}.  

2)   The time required for crater excavation.
 
3)   A time for penetration of the projectile in the granular medium by a distance equal to the crater depth.   

We compare the time for a sound wave to propagate back and forth across the projectile (listed
as $t_{D,prop}$ in Table \ref{tab:imp}) to the pulse durations which are listed as $\Delta t_{a}$ in 
Table \ref{tab:exps}.  The sound propagation times across the projectile 
are 0.7 and 0.01 ms for our two projectiles and are a poor match to our pulse durations.    
The time is shorter in the hard glass marble projectile than the softer rubber ball. 
\citet{yasui15} found that the sound crossing time was only a few microseconds for their projectiles 
and too small to match their pulse durations.  We concur with \citet{yasui15} that the sound travel
time across the projectile does not set the seismic pulse duration.

Using a high speed camera,  \citet{yasui15} estimated that their craters formed in 100 to 200 ms, which 
exceeds their pulse widths.   High speed videos of our ejecta curtains give a similar timescale.
We concur with \citet{yasui15} that the crater excavation time does not set the seismic pulse duration.

The normal impact experiments into granular media by \citet{goldman08} and \citet{murdoch17} have impact velocities that are similar to ours.  
These studies measured a stopping time, 70--200 ms, 
which is a time for the projectile to come to rest. 
With a high speed camera video taken at 1069 fps
we measured that the time for the projectile to come to rest in the MR5 experiment
was about $t_{stop} \sim 30$ ms.   Our videos are not as sensitive to low velocities 
as an accelerometer embedded inside a projectile -- this difference might account
for the longer stopping times measured from other normal impact experiments 
into granular media compared to ours.
Our estimated stopping time exceeds our pulse duration by an order of magnitude. 

The stopping time, or time it takes the projectile to come to rest, differs from 
the decay time for initial deceleration which might be more relevant for seismic pulse excitation.
In empirical models for projectile deceleration during a normal impact
(\citealt{tsimring05,katsuragi07,goldman08,katsuragi13}),
the force on the projectile is the sum of a drag-like term that is proportional to the square of
the projectile velocity 
and a depth dependent and velocity independent term.
When the velocity is high, the drag-like term dominates, giving equation of motion 
\begin{equation}
m_p \frac{dv_p}{dt}   \approx -\frac{1}{2} C_D \rho_s \pi R_p^2 v_p^2,   \label{eqn:emp}
\end{equation}
where $v_p$ is the speed of the projectile and $C_D$ is a drag coefficient
that is of order unity \citep{katsuragi13}. 
The regime where the velocity independent forces are neglected is called the inertial regime
and the velocity squared dependence for the force in this regime is well supported by normal impact  experiments into granular media 
 \citep{goldman08,murdoch17}. 
For a spherical projectile, Equation \ref{eqn:emp} has solution for depth, speed and acceleration 
\begin{align}
-z_p(t) & =  \frac{1}{ \alpha_p} \ln (v_{imp} \alpha_p t + 1) \nonumber \\
v_p(t) & = \frac{v_{imp}}{v_{imp} \alpha_p t + 1}   \nonumber \\
a_p(t) & = \frac{v_{imp}^2 \alpha_p}{(v_{imp} \alpha_p t + 1)^2}, \label{eqn:accel}
\end{align}
consistent with Equation 8 by \citet{yasui15}.
Here $v_{imp}, v_p$ and $t$ are assumed to be positive and 
the inverse length-scale 
\begin{equation}
\alpha_p =  \frac{3C_D }{8R_p} \frac{\rho_s}{\rho_p}  \label{eqn:alpha_p}
\end{equation}
is equivalent to the drag parameter $1/d_1$ by \citet{katsuragi13}.
For our experiments with $C_D=1$ we find $\alpha_p = 14,15 $ m$^{-1}$ for ML5, SL5
experiments respectively.    Deceleration is characterized by a time 
\begin{equation}
t_{decel} = \frac{1}{\alpha_p v_{imp}}, 
\end{equation}
which for our experiments is $ t_{decel} \sim 15$ ms. This is too long to 
 match our pulse durations $\Delta t_v$.
\citet{yasui15} mitigated this problem by using the same type of empirical model 
but computing the time for the projectile 
to cross a distance equal to the crater depth, giving a time they denoted a penetration time. 
 
As projectile stopping or deceleration times give a poor match to our pulse durations,  
we consider additional processes that might account for them. 
Impacts into granular media such as sand obey scaling laws that are independent of material
strength and so are in the gravity regime \citep{holsapple93}.  In the gravity regime 
the crater volume and radius are approximately set by scaling laws that only depend on density
ratio $\pi_4$ and the $\pi_2$ parameter that is a function of the Froude number.  
We expect that the crater radius is set by the impactor size and 
speed and the substrate density.  Additional properties of the medium, such as pulse propagation velocity, $v_P$, 
could influence other characteristics of the impact such as the pulse duration. 

A seismic source can have a cutoff frequency in its spectrum that is set by
the size of the seismic source region \citep{aki}.   A related time-scale would be
the seismic source size divided by the seismic wave travel speed. 
For periods shorter than the seismic source time, seismic waves emitted from the near and far sides of the source would interfere, and this would give a minimum duration in a seismic pulse.
The crater itself is a candidate for the size of the seismic source and this 
gives a crater seismic source time-scale
\begin{equation} 
	t_{Rss} = \frac{R_{cr}}{v_P}  \label{eqn:tRss}
\end{equation}
where $v_P$ is the speed that the pulse travels through the medium.

Using the crater radii listed in Table \ref{tab:exps} and the pulse travel speed 
(listed in Table \ref{tab:pressure} and discussed in Section \ref{sec:speed}),
we estimate crater seismic source times
of about 1.0 and 0.75 ms, respectively for the ML5 and SL5 experiments.  These seismic
source times are also summarized in Table \ref{tab:duration}. 
The crater seismic source times are similar to the duration of the pulses seen in the first accelerometer 
$\Delta t_{a} \approx 1.0$ and 0.5 ms, respectively for the same experiments.
The crater radius in the sand experiment is smaller than those in the millet experiments, supporting
the association of the $t_{Rss}$ with the seismic pulse duration.

\begin{figure}[ht] \centering 
\includegraphics[width = 3 truein, trim = 20 20 0 0, clip]{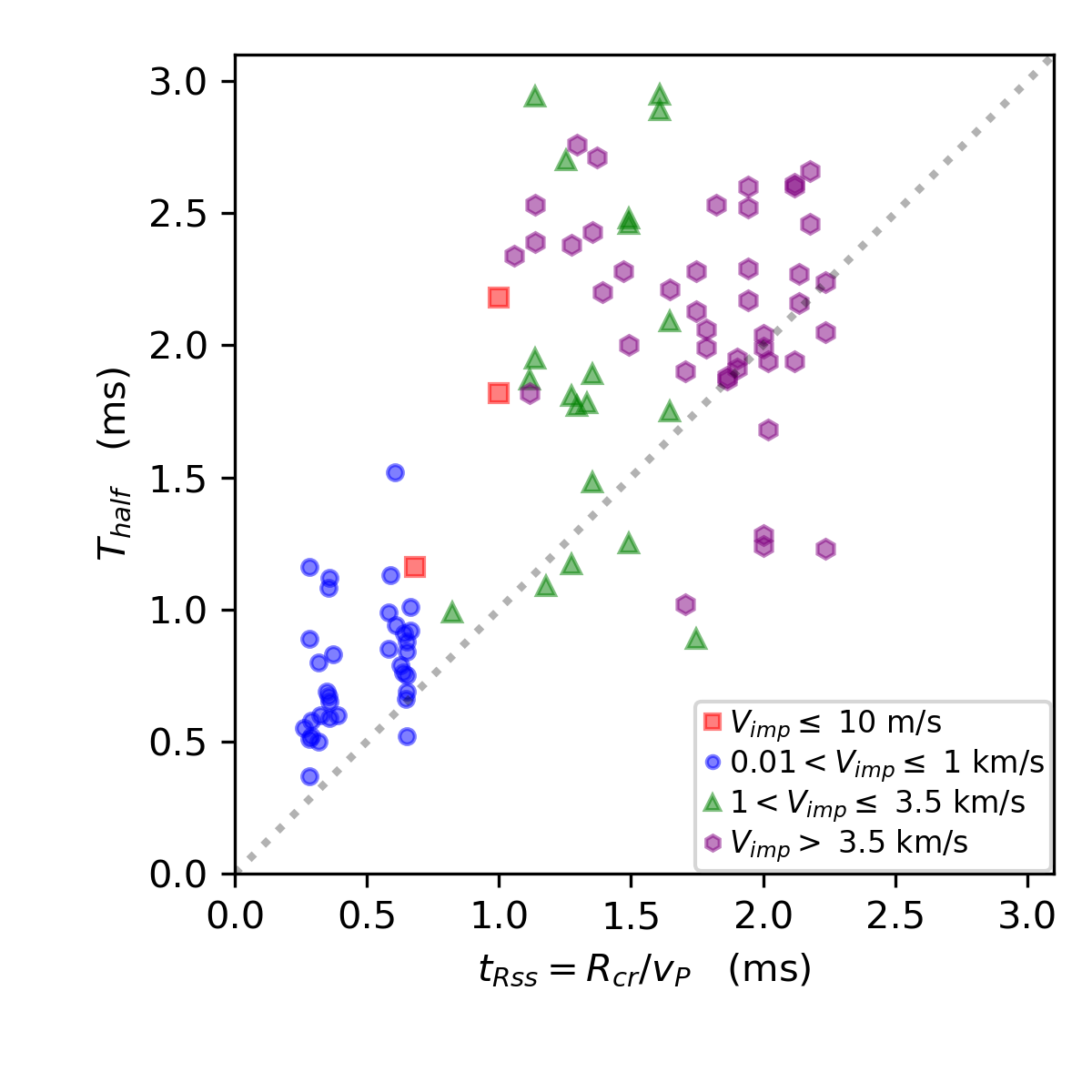}
\caption{Pulse durations are plotted against the seismic source time.
Measurements by \citet{yasui15} are shown with blue circles and these correspond
to experiments with impact velocity about 100 m/s.
Measurements by \citet{Matsue_2020} are shown with blue circles, green triangles or purple hexagons,
depending upon the impact velocity of the experiment. 
Our experiments are shown with red squares. 
We only plot measurements from accelerometers that are between 1 and 2.2 crater radii
of the impact site. 
\label{fig:TT}}
\end{figure}

Is the crater seismic source time similar to 
the pulse duration in the higher velocity impact experiments into granular media by \citet{yasui15} and \citet{Matsue_2020}?  
Using the crater radii listed in Table 1 by  \citet{yasui15}
and Tables 3, 4 and 5 by \citet{Matsue_2020}, we estimate the seismic source times  
and plot them as a function of the pulse duration $T_{\rm half}$ which is the duration of
the first positive portion of the acceleration pulse.  
The results are shown in Figure \ref{fig:TT}.
The seismic source time is computed as $R_{cr}/v_P$ (equation \ref{eqn:tRss}) using
 the pulse travel speed $v_P$ measured by these experiments.  
For the experiments into glass beads by  \citet{yasui15} 
we use pulse travel speed $v_P = 110 $ m/s, consistent
with their measurement, and
we use $v_P = 50$ m/s for the into quartz sand by \citet{Matsue_2020}, consistent
with their measurement.
We also plot points for our experiments on Figure \ref{fig:TT}.   
We multiply our pulse durations $\Delta t_a$
by two, as our pulse FWHMs are about half the duration of the $T_{\rm half}$ parameter
measured by \citet{yasui15} and \citet{Matsue_2020}.
On Figure \ref{fig:TT} 
to remove any possible sensitivity to travel distance, 
we only plot points from accelerometers that have distances $1 < r/R_{cr} <2.2$.
A dotted grey line on the figure shows $T_{\rm half} = t_{Rss}$, illustrating
that to order of magnitude the two times are similar in size.  In Figure \ref{fig:TT},  
the different impact velocities are shown with different colored and shaped points. 
Figure \ref{fig:TT} shows that over a wide range in impact velocity,  
pulse duration is correlated with the seismic source time.
We did not exclude measurements from experiments with high density impactors, so
the similarity between seismic source timescale
and pulse duration holds for different density impactors.  The relation is
 not strongly sensitive to impactor density. 

Is the crater seismic source time consistent with pressure pulse duration measured for impacts into solids? 
The pulse widths for the impact experiments discussed by \citet{Guldermeister_2017}
into sandstone and quartzite have durations of about 5 $\mu$s.  Their crater radii were
a few cm and the pulse propagation speeds a few km/s.  We find that the ratio of crater radius
to pulse propagation speed is also similar in size to the pulse durations for the impact experiments 
by \citet{Guldermeister_2017}.

As the projectile penetrates the granular medium, pressure in the medium below the projectile 
increases.   The duration of the pulse driven into the medium could be related to the mechanism for release of this pressure, in analogy to the role of the time for a shock
 to propagate forward and a rarefaction wave to propagate backward across the projectile
 following a high velocity impact \citep{melosh89}.  
The pressure pulse propagates through the medium at the speed $v_P$.  The relevant distance for pressure release 
 would be the crater radius.  
This gives a propagation time $R_{cr}/v_P$ which is equal to the crater seismic source
time of Equation \ref{eqn:tRss}.  This heuristic physical explanation for the seismic pulse duration might account for
the similarity between pulse duration and the crater seismic source time-scale in our and other experiments.



Could pulse broadening of an initially narrow pulse (e.g., \citealt{Hostler_2005,Owens_2011,Langlois_2015,Zhai_2020}) 
and associated with scattering account for our pulse durations?  
For pulses that are initially very short duration and propagate through a granular medium, 
\citet{Langlois_2015} introduced a normalized pulse duration 
\begin{equation} 
	W \equiv \frac{\Delta t}{ t_{prop} }  \label{eqn:W}
\end{equation}
where $\Delta t$ is the pulse duration, as seen in pressure or velocity, and $t_{prop}$ is the pulse travel or propagation time.
Because we are working with pulse velocities and accelerations, we define similar 
normalized pulse widths
\begin{align}
W_a  \equiv \frac{\Delta t_a}{t_{a,prop} },  \qquad \qquad 
W_v  \equiv \frac{\Delta t_v}{t_{v,prop} }. \label{eqn:W_av}
\end{align}
Normalized pulse widths for the accelerometers nearest
the impact site are also listed in Table \ref{tab:duration} for the MR5, ML5 and SL5 experiments.
\citet{Langlois_2015} found that the normalized pulse width 
\begin{equation}
	W \approx C_W \sqrt{\frac{d_g}{r}}  \label{eqn:C_W}
\end{equation}   
where $d_g$ is the grain diameter, $r$ is the travel distance and dimensionless coefficient $C_W \sim 1$. 

Using a mean grain diameter of 3 mm for millet and 0.3 mm for the fine sand
we estimate $\sqrt{d_g/r} \sim 0.2$ for the first accelerometer in the ML5 experiment
and $\sqrt{d_g/r} \sim 0.06$ for the first accelerometer in the SL5 experiment.
However, normalized pulse widths in velocity are $W_v \sim 1$.   
Only if the coefficient $C_W$ is 5 to 20 and significantly greater than 1 could
the diffusive model by \citet{Langlois_2015} match the normalized velocity pulse widths,
assuming that pulse widths were initially shorter than a ms.
The rate of pulse broadening could be sensitive to pulse
peak pressure, as suggested by the pulse pressure dependent attenuation rate 
seen in experiments of strong pulses \citep{vandenWildenberg_2013}, so a higher
effective value of $C_W$ is possible. 


In summary, a time based on the size of the crater and the speed that pulses
travel through the granular medium is a possible time-scale that could account for the duration of the pulses we see in our experiments and in higher velocity experiments.  The shorter duration pulses in sand suggest that pulse duration could also be sensitive to grain size, as suggested by models of pulse broadening \citet{Langlois_2015}.    Our pulse durations could be consistent with an initially short duration pulse (less than 1 ms)
and subsequent broadening described by Equation \ref{eqn:C_W}, but only if the scaling coefficient $C_W$ is about 10 and an order of magnitude larger than measured by \citet{Langlois_2015}.

\subsection{Pulse broadening}
\label{sec:broad}

\begin{figure}[htbp] \centering 
\if \ispreprint1
\includegraphics[width = 3.3 truein, trim = 0 0 0 0, clip]{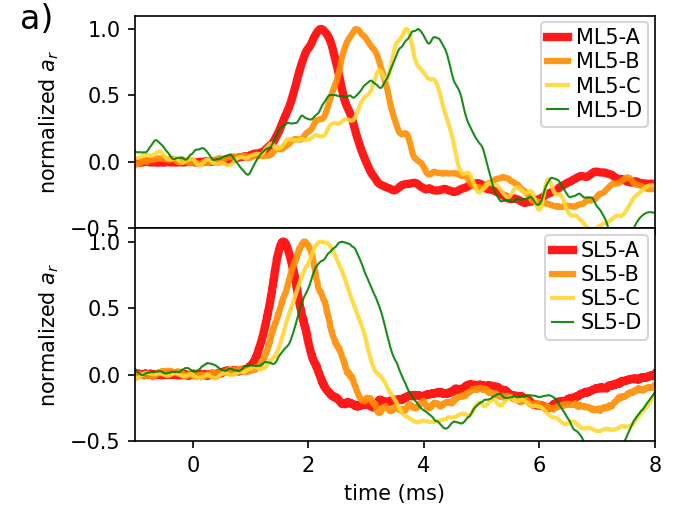}
\includegraphics[width = 3.3 truein, trim = 0 0 0 0, clip]{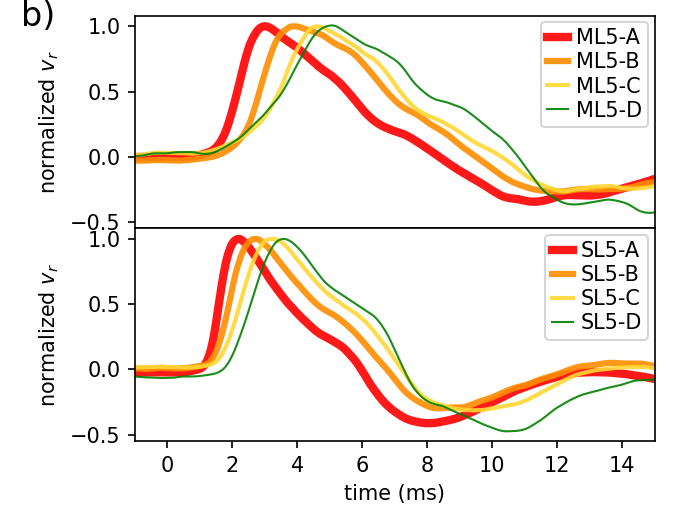}
\else
\includegraphics[width = 4.0 truein, trim = 0 0 0 0, clip]{nov27ar_norm.png}
\includegraphics[width = 4.0 truein, trim = 0 0 0 0, clip]{nov27vr_norm.png}
\fi
\caption{a) The radial component of acceleration is
shown as a function of time for the four accelerometers nearest the impact site
from the ML5 experiment into millet in the top panel and in the SL5 experiment into sand in the bottom panel.
Here A, B, C, D 
refer to the accelerometers in order of distance from the site of impact. 
The signals have been scaled so that they have the same peak heights.
Pulses broaden and become smoother as they travel, and   
the pulses in the millet experiment are broader than those in the sand experiment.
Smoothing is particularly noticeable on the rising
side of the pulses. 
b) Similar to a) except we show velocities.  
\label{fig:norm}
}
\end{figure}

\begin{figure}[htbp] \centering 
\if \ispreprint1
\includegraphics[width = 3.3 truein, trim = 0 0 0 0, clip]{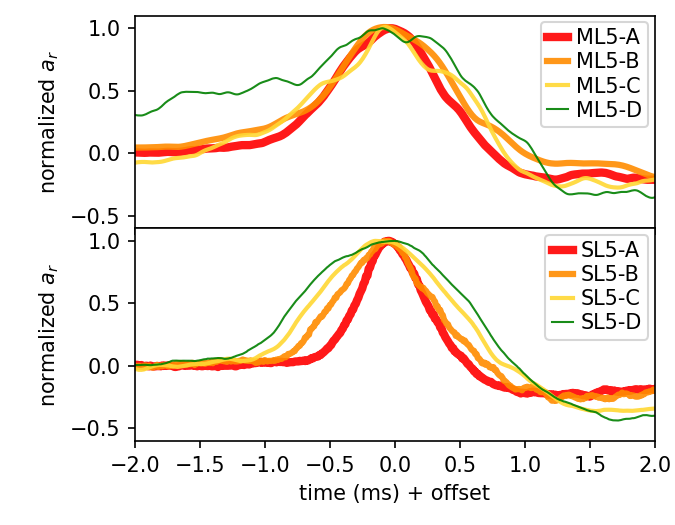}
\else
\includegraphics[width = 4.3 truein, trim = 0 0 0 0, clip]{nov27broad.png}
\fi
\caption{Similar to Figure \ref{fig:norm}a showing accelerations except times have been shifted so that the peaks occur at a time of about zero.
\label{fig:broad}
}
\end{figure}

To best study variations in pulse shape, we look at the strongest signals which are 
those nearest the site of impact. 
We focus on the radial accelerations and velocity from the four accelerometers nearest the site
of impact in the ML5 experiment into millet and the SL5  experiment into sand.  
These two experiments have the same coordinate positions for the accelerometers
and their signals were previously shown in 
Figures \ref{fig:ML5} and \ref{fig:SL5}.  
In Figure \ref{fig:norm} we show radial acceleration and velocity signals normalized
so that the peaks have the same height. Figure \ref{fig:broad} is similar but 
 the signals have been shifted in time so that the peaks are near a time of zero.
%
Figure \ref{fig:norm} and  \ref{fig:broad} shows that the pulses are broader in millet than in sand
and that the pulses broaden and become smoother as they travel through the granular medium.

\begin{figure}[htbp] \centering 
\if \ispreprint1
\includegraphics[width = 3.3 truein, trim = 0 0 0 0, clip]{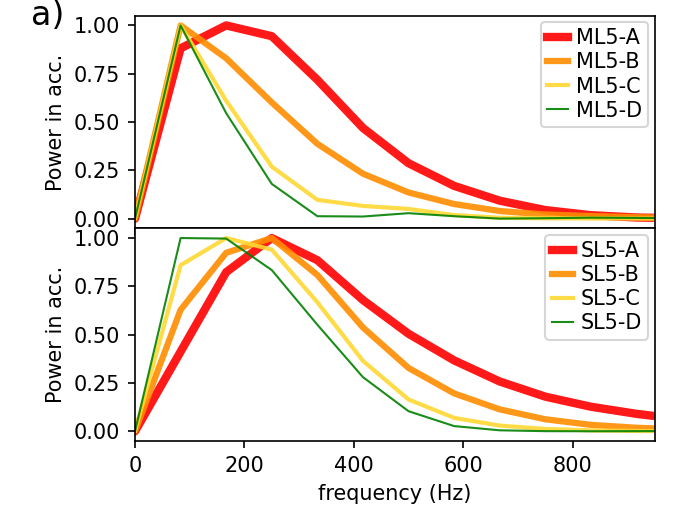}
\includegraphics[width = 3.3 truein, trim = 0 0 0 0, clip]{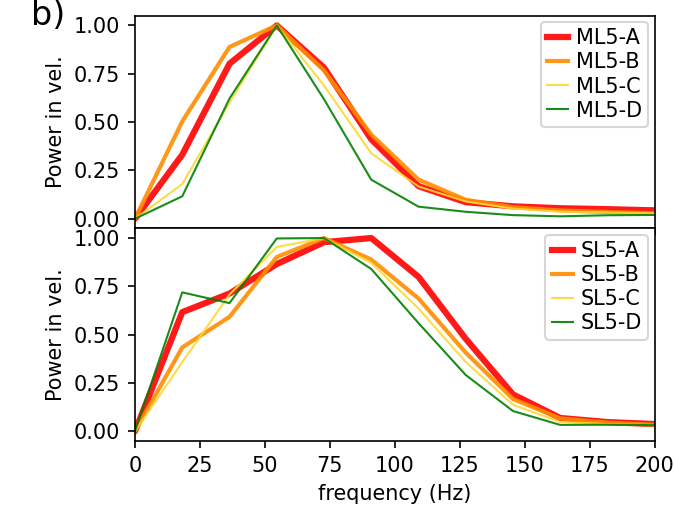}
\else
\includegraphics[width = 4.3 truein, trim = 0 0 0 0, clip]{nov27power_a.png}
\includegraphics[width = 4.3 truein, trim = 0 0 0 0, clip]{nov27power_v.png}
\fi
\caption{a) Power spectra of the acceleration pulses
from the four accelerometers nearest the point of impact from the ML5 (top panel)
and SL5 experiments (bottom panel).
b) Similar to a) except showing power spectra of the velocity pulses. 
\label{fig:power}
}
\end{figure}

In Figure \ref{fig:power} we show power spectra of the acceleration signals shown in Figure \ref{fig:norm} 
as a function of frequency.   The signals are multiplied by a Hanning window function and the mean subtracted prior to computing the Fourier transform. 
We use a window that is 12 ms long for the accelerations and 35 ms long
for the velocities. 
The power spectra have been normalized so that their peak is 1.
The spectrum has zero power at zero frequency because the mean signal value was subtracted. 
Figure \ref{fig:power}a shows that the accelerometers nearest the site of impact peak 
have more power at higher frequencies than those more distant from the site of impact. 

Pulse broadening is most clearly seen in the sand experiment 
(see bottom panels of Figures \ref{fig:norm}a, and \ref{fig:broad}) in part because the pulse 
is initially shorter.  However, smoothing of the pulse shape is most clearly seen early, where
the pulse accelerations and velocities rise in both sand and millet experiments 
(see Figures \ref{fig:norm}a,b).   
If attenuation is dependent upon frequency, pulses would
become smoother as they travel. This is consistent with the power spectra we show in Figure \ref{fig:power}
that show that the pulses preferentially lose power at higher frequencies as they travel. 
We attribute steep slopes where velocities drop in the accelerometers most
distant from thee impact site to a reflection from the tub rim. 

We discuss the pulse duration regime for our experiments.   Our pulse durations are about 
a ms and correspond to a spatial width $\sim $ 50 mm (estimated using $v_P$),  and this 
corresponds to a wavelength
of about $\lambda \sim 100$ mm.   Our grains have diameters of 3 and 0.3 mm respectively for millet and sand, respectively.    This gives ratio of wavelength to grain diameter $\lambda/d \sim 30$ and 300 for the millet and sand experiments, respectively. 
The experiments by \citet{Langlois_2015} in glass beads had grain diameters ranging from 0.22 to 5 mm and the pulse travel speed was 800 m/s. 
Their pulse durations ranged from about 5 to 42 $\mu s$ (following their Figure 5a) giving spatial 
pulse widths of 4 to 34 mm, corresponding to wavelengths of $\lambda \sim $ 8 to 70 mm.
This would give $\lambda/d$ ranging from  about 2 to 300.  Because pulse durations were in the regime $\lambda/d >1$, \citet{Langlois_2015} proposed that the attenuation was due to scattering caused by variations in grain stiffness, rather than due to variations in travel
 times along different force contact chains \citep{Owens_2011}.   \citet{Hostler_2005} measured 
 pulse broadening in a regime where pulse width was about $10^3$ times the pulse travel time across
a single grain giving $\lambda/d  \sim 10^3$.    
Thus at least three sets of experiments have measured pulse broadening in the regime  $\lambda/d >1$.
 
The scaling observed by between pulse width and travel distance by  \citet{Langlois_2015} obeyed scaling that they interpreted as due to attenuation rather than dispersion. 
Broadening and smoothing of the pulse shape is likely to be connected to dissipation mechanisms.
In contrast,  a dispersive mechanism need not be associated with dissipation. 
Energy can be dissipated via several mechanisms which can operate at $\lambda/d>1$, including frictional and inelastic particle interactions (discussed by \citealt{Hostler_2005}),
particle rearrangements (discussed by \citealt{Zhai_2020}),  scattering through the particle contact network \citep{Owens_2011} and scattering
due to variations in particle stiffness \citep{Langlois_2015}, variations in the packing fraction or porosity and 
variations in the connectivity of the particle contact network. 


\section{Pulse smoothing and attenuation}
\label{sec:diff}

Figures \ref{fig:ML5}, \ref{fig:SL5}, \ref{fig:MR5}, \ref{fig:norm}, \ref{fig:broad} and \ref{fig:power} 
illustrate that the pulses in our experiments broaden and become
smoother as they travel  through the granular medium.  The pulses do not break up
into a series of high frequency waves followed by low frequency ones, or vice versa,  as
would be expected for dispersive model for wave propagation.   We primarily
observe attenuation, smoothing and broadening, which are characteristics of diffusion 
and not of dispersion.   

Granular systems can display both solid-like and fluid-like behavior (e.g., \citealt{GDRMiDi_2004,Forterre_2008}). We 
 introduce both elastic and hydrodynamic continuum models for wave propagation. 
%
For propagation of elastic waves in one dimension in an isotropic medium,  
momentum conservation can be written as 
\begin{equation}
\rho \frac{\partial u }{\partial t}     = -\frac{\partial \sigma } {\partial x} 
\end{equation} 
where the velocity field $u(x,t) = \frac{\partial \delta} {\partial t} $,   with 
$\delta(x,t)$ the displacement field,  and $\sigma$ is one component of the stress tensor.  
With stress linearly proportional to the strain (which depends on the gradient of the
displacement field)  and in
the low amplitude limit, 
 a wave equation for propagation of longitudinal waves is derived
\begin{equation}
\frac{\partial^2 \delta}{\partial t^2} = v_P^2  \frac{\partial^2 \delta}{\partial x^2},  \label{eqn:wave1}
\end{equation}
(e.g., \citealt{aki}, section 2).
This gives a dispersion relation  $ \omega^2 = v_P^2 k^2 $ where $\omega$ is
the angular frequency of a traveling sine wave that has wave number $k$.
A more general model for the dispersion relation would give a complex and non-linear function that 
is non-linear.   If $\omega(k)$ or $k(\omega)$ has a complex component, this can be interpreted
in terms of a wavelength or frequency dependent decay rate for the amplitude.  
This is commonly described as attenuation.  
If $\omega(k)$ is real but non-linear then the system is described as dispersive. 
Dispersion naturally arises 
if the Taylor expansion of the elastic stress depends on the second or higher order spatial derivatives of the displacement field.   In this case the model is said to be `anelastic'.

If dispersion and attenuation are low and pulses propagate  
in one direction to positive $x$ then equation \ref{eqn:wave1} 
is consistent with $ \frac{\partial u}{\partial t}  +v_P \frac{\partial u}{\partial x}=0$ which is known as the advection equation.   
The same relation can be derived for hydrodynamics in the low
amplitude limit 
using Euler's equation, conservation of mass and an equation of state,
with $v_P$ equivalent to the velocity of sound.  Conservation of mass to first order in perturbation amplitudes, 
the equation of state and a nearly wavelike solution gives 
 $\frac{1}{\rho} \frac{dp}{dx} \approx v_P \frac{du}{dx}$  
where $p$ is pressure.  Neglecting the non-linear inertial term,  the Navier Stokes equation becomes
\begin{equation}
\frac{\partial u}{ \partial t} = -v_P \frac{ \partial u}{ \partial x} +D \frac{\partial^2 u }{\partial x^2}. \label{eqn:wave2}
\end{equation}
Here $D = \nu/v_P$ depends on the kinematic viscosity $\nu$.   
The Navier Stokes equation is also derived via momentum
conservation with stress tensor dependent upon pressure, ram pressure, viscosity and 
the velocity gradient.  
Equation \ref{eqn:wave2} is an advection-diffusion equation.  The integral $\int u(x,t) dx$ is a conserved quantity, so density variations need not be taken into account to maintain 
conservation of momentum. 
A solution to equation \ref{eqn:wave2} that is initially a delta function is 
\begin{equation}
u(x,t)  = \frac{1}{\sqrt{4 Dt}} e^{-\frac{(x - v_P t)^2}{4 D t} }. \label{eqn:Greens}
\end{equation}
The dispersion relation for equation \ref{eqn:wave2} is
$\omega = v_P k + i D k^2$ and the viscous or diffusive behavior causes attenuation.

Granular flows often exhibit a dependence on the shear rate, which gives them a viscous-like behavior
\citep{Forterre_2008}.  
Viscous or diffusive behavior in the context of elastic waves arises naturally
if the stress tensor is dependent on the strain rate, which depends 
on the time derivative of the displacement field.     
The stress can gain a term dependent on $\frac{\partial^2 \delta}{\partial x\partial t}$, giving
an additional term $\propto \frac{\partial^3 \delta} {\partial x^2\partial t} = \frac{\partial^2 u}{\partial x^2} $
in the wave equation (equation \ref{eqn:wave1}), which resembles the diffusive term in equation \ref{eqn:wave2}.  In this case, the model is said to be `visco-elastic'. 
 
Because equation \ref{eqn:wave2} contains a diffusion term, short wavelength structure
attenuates more quickly than those at longer wavelengths.  A velocity pulse will become
smoother as it travels and an initially narrow pulse will broaden.  
The association of equation \ref{eqn:wave2} with conservation of momentum motivates
using the velocity field as a key variable. 

Because our experiments primarily show attenuation and smoothing, rather than dispersive behavior, 
we adopt a diffusive model, with velocity as key variable  to model the propagation of our pulses. 
Our experiments predominantly show a single pulse that rapidly attenuates 
as it travels, confirming results from prior experiments of impacts into granular media
\citep{mcgarr69,yasui15,Matsue_2020}.   This motivates describing the pulse with
two parameters: an amplitude and a duration. 
An advantage of an advection-diffusion model for pulse propagation into a half sphere is that the rate
that the pulse duration grows 
is directly related to the pulse amplitude decay rate.
This gives a simple model
that predicts how pulse amplitude, pulse width and energy vary with propagation distance.
We can compare the resulting scaling relations to the dependence of 
pulse width and peak amplitudes on distance from impact site in our experiments.

We first show that the model for pulse broadening proposed by \citet{Langlois_2015} 
is consistent with an advective-diffusion model for velocity
propagation.  
We do this showing that  
 their scaling law (Equation \ref{eqn:C_W}), relating pulse width to travel distance,  can be derived via 
 a diffusive model characterized with a diffusion coefficient $D$.
The propagation distance $r$ is related to the pulse travel time 
 via $r = v_P t_{prop}$ where $v_P$ is the pulse travel speed.   
We define $\Delta r = \Delta t v_P$ as the spatial pulse width.
 Diffusive broadening of a delta function at $t=0$ (as in equation \ref{eqn:Greens}) gives pulse width  
$\Delta r \sim \sqrt{D t_{prop}}$ as a function of propagation time $t_{prop}$. 
This spatial width corresponds to a pulse duration 
\begin{equation}
\Delta t \approx   \frac{ \sqrt{D t_{prop}}}{v_P}  \approx \sqrt{ D r/v_P^3}.
\label{eqn:dtr}
\end{equation}

Using Equations \ref{eqn:W} and \ref{eqn:dtr},
the normalized pulse duration
\begin{equation}
W \sim \sqrt{\frac{D}{r v_P}}.
 \label{eqn:W2}
\end{equation}
Comparison of this equation with the 
scaling relation by \citet{Langlois_2015} in Equation \ref{eqn:C_W} implies that this relation 
is consistent with  
 diffusion coefficient that depends on pulse propagation speed and grain size 
\begin{equation}
D \sim C_W^2 d_g v_P. \label{eqn:D}
\end{equation}
We have illustrated that the scaling relation  (Equation \ref{eqn:C_W}) 
by \citet{Langlois_2015} can be derived via a diffusive model for pulse propagation of an
initially narrow pulse and 
with diffusion coefficient given by Equation \ref{eqn:D}.
\citet{Tell_2020} also adopted a diffusive model for scattering
and their experiments were also consistent with a diffusion coefficient
in the form of Equation \ref{eqn:D} with  $C_W \sim 1$ (see their Section VI).

If the pulse is not initially a delta function then diffusive broadening gives a pulse
duration 
$\Delta t \approx \sqrt{\Delta t_0^2 +  D r/v_P^3}$ 
where $\Delta t_0^2 $ is the duration at $t=0$ and at $r=0$.
Instead of writing pulse duration in terms of that at $t=0$ we scale
from the pulse duration for the pulse at a distance of the crater radius.
The pulse duration for $r > R_{cr}$ 
\begin{equation}
\Delta t (r)\approx    \sqrt{ \frac{D (r- R_{cr})} {v_P^3 }  + \Delta t_{Rc}^2 } 
,
\label{eqn:dtr_0}
\end{equation}
where $\Delta t_{Rc}$ is pulse width at $r = R_{cr}$.

\begin{figure}[htbp] \centering 
\if \ispreprint1 
\includegraphics[width = 3.3 truein, trim = 20 10 0 0,clip]{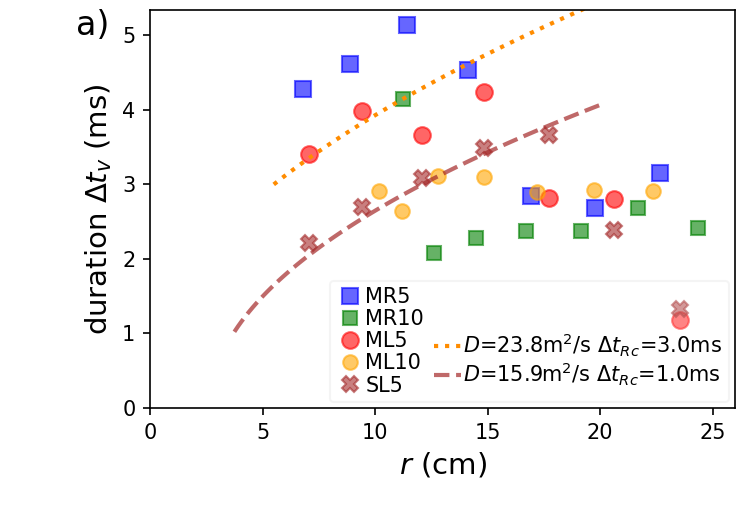}
\includegraphics[width = 3.3 truein, trim = 20 10 0 0,clip]{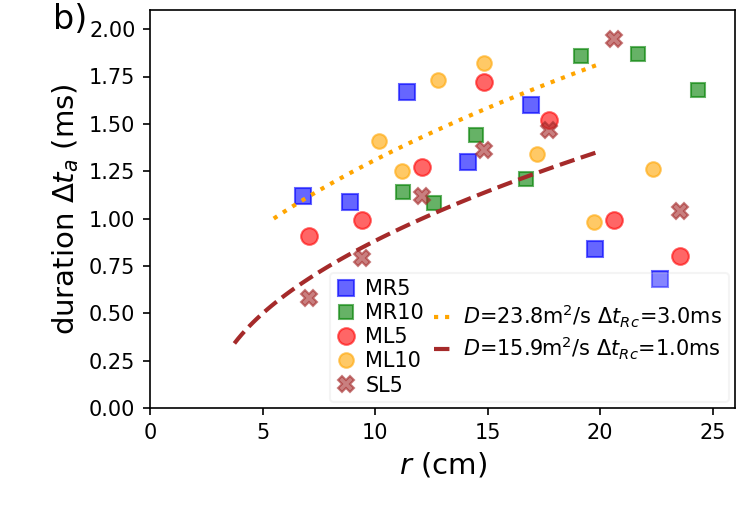}
\else
\includegraphics[width = 4.3 truein, trim = 20 10 0 0,clip]{nov27fwhm_v.png}
\includegraphics[width = 4.3 truein, trim = 20 10 0 0,clip]{nov27fwhm_a.png}
\fi
\caption{a) Pulse duration, the FWHM, $\Delta t_v$,  in ms 
as a function of travel distance for the MR5, ML5, ML10, MR10 and SL5 experiments. 
Brown dashed and orange dotted lines are computed using Equation \ref{eqn:dtr_0},  
diffusion coefficients $D$ and duration $\Delta t_{Rc}$ at $r=R_{cr}$ listed in the key.    
b) Similar to a) except showing $\Delta t_a$.  
\label{fig:fwhm}
}
\end{figure}

In Figure \ref{fig:fwhm}a we plot pulse duration $\Delta t_v$ (the full width half max
of the first positive portion of the radial velocity pulse) for the MR5, ML5, ML10, SL5 and SL10 experiments. 
On this plot, the dashed brown and dotted orange lines show pulse durations estimated with Equation \ref{eqn:dtr_0} and with diffusion coefficient $D$ and duration $\Delta t_{Rc}$ listed in the key.  
The orange dotted line uses $R_{cr}$ for the millet experiments and the brown
dashed lines uses $R_{cr}$ for the sand experiment. 
Using Equation \ref{eqn:D} and a mean size for the millet grains ($d_g = 3$ mm),  the diffusion coefficient for the dotted orange line is consistent with scaling coefficient $C_W = 12$.
Using the mean sand grain size (0.3 mm), the dashed brown line gives $C_W = 30$,    
The diffusion coefficient is lower in the sand, as would be expected if the broadening rate were sensitive to grain size.  
However, the mean sand grain 
size is about an order of magnitude smaller than the size of the millet grains and Equation \ref{eqn:D}
predicts $D \propto d_g$.  
The $C_W$ values are about an order of magnitude larger than expected (in comparison
to experimental measurements by \citealt{Langlois_2015,Tell_2020}), as we previously estimated
in Section \ref{sec:duration}.
The initial durations we use for
the dotted orange and dashed brown lines have $\Delta t_{Rc}$ at $r = R_{Cr}$ that 
are 3 times and 1.5 times $R_{cr}/v_P$ for the millet and sand
experiments, respectively. As discussed in Section \ref{sec:duration}, these initial durations are similar to the seismic source times. 
Figure \ref{fig:fwhm}b is similar to Figure \ref{fig:fwhm}a except we show $\Delta t_a$ instead
of $\Delta t_v$.   Pulse broadening is also seen Figure \ref{fig:fwhm}b.

Figure \ref{fig:fwhm} shows that pulses initially broaden with an estimated diffusion coefficient 
that exceeds that which we would have estimated using relations by  \citet{Langlois_2015,Tell_2020}, with $C_W \sim 1$, 
by about an order of magnitude.  Past about 15 cm from the impact site, pulses are weaker and the widths could be affected
by reflections, so  we are not concerned by the low durations measured from the accelerometers 
more distant from the impact site. 

The curves on Figure \ref{fig:fwhm} illustrate that the diffusion coefficient is sufficiently large that 
even a few crater radii away from impact, the diffusion term in Equation \ref{eqn:dtr_0} dominates the pulse duration
and the initial pulse duration is less relevant.  In other words, Equation \ref{eqn:dtr_0}
can be approximated by Equation \ref{eqn:dtr}.
The simpler power-law form of Equation \ref{eqn:dtr} facilitates predicting attenuation, so we will
use it in the following sections. 

\subsection{Pulse amplitudes}
\label{sec:amp}

With pulse broadening due to diffusion of a pulse's velocity, what attenuation is implied?
We consider a pulse that propagates radially into a half sphere
with peak amplitude in velocity $v_{pk}(r)$ which is a function of  propagation distance.
If the pulse propagates radially from the origin and isotropically, then 
momentum conservation implies that 
\begin{equation}
\rho_s v_{pk}(r) \Delta t_v v_P 2 \pi r^2  = {\rm constant}. \label{eqn:cc}
 \end{equation} 
This follows by integrating the momentum flux as a function of time in a pulse that passes radius $r$ and in 
a small region of solid angle. 
Here $\Delta t_v$ is the width of the velocity or pressure pulse. In Equation \ref{eqn:cc} we neglect the angular dependence of pulse amplitude, however this equation could be modified to depend on the spherical coordinate polar angle from the surface normal. 
Equation \ref{eqn:cc} and using Equation \ref{eqn:dtr} for $\Delta t_v$ 
implies that the peak velocity 
\begin{equation} 
	v_{pk} (r) 
	\propto r^{-\frac{5}{2}} v_P^{\frac{1}{2} }D^{-\frac{1}{2}}   .
     \label{eqn:vpkr}
\end{equation}

What do we expect for the radial scaling of the magnitude of the peak acceleration?  
The peak acceleration  $a_{pk} \sim \frac{v_{pk}}{\Delta t_v}$.
Using Equation \ref{eqn:vpkr} for the peak velocity and Equation \ref{eqn:dtr} for $\Delta t_v$ 
\begin{align}
a_{pk}(r) 
& \propto r^{-3} v_P^{2} D^{-1}  . \label{eqn:apkr}
\end{align}

The peak particle displacement in the pressure pulse we estimate from Equations \ref{eqn:vpkr}
and \ref{eqn:dtr} 
\begin{align}
\delta_{pk}(r)  \sim v_{pk} \Delta t_v  \propto r^{-2} v_P^{-1}
. \label{eqn:dpkr}
\end{align}


Above we have estimated the radial scaling of peak acceleration and velocity.
We now use the momentum of the projectile to estimate the constants of proportionality
for Equations \ref{eqn:vpkr} and \ref{eqn:apkr}.
Using Equation \ref{eqn:accel}, 
the projectile deceleration at the moment of impact is
\begin{equation}
a_{p,max} \approx - \alpha_p v_{imp}^2. \label{eqn:apmax}
\end{equation}
If the projectile deceleration is due to the launch of a pressure pulse within the medium, then 
momentum conservation can be used to estimate the size of the pressure pulse.
At the crater radius we assume that 
\begin{equation}
b_{\rm eff} m_p \frac{dv_p}{dt} \sim \rho_s \pi R_{cr}^2 v_{pk}(R_{cr}) v_P  \label{eqn:vpert}
\end{equation}
where on the right side the momentum flux from a pulse that travels from the projectile surface 
with velocity $v_P$ and with a peak radial velocity perturbation $v_{pk} (R_{cr})$ at distance 
$r = R_{cr}$ from the impact site is matched 
to the deceleration of the projectile which is on the left.  
We include a dimensionless factor $b_{\rm eff}$, similar to the factor known as the 
momentum transfer efficiency or $\beta$ parameter 
\citep{housen11,Holsapple_2012,Jutzi_2014}, which is
the total momentum change of an asteroid resulting from an impact, divided by the projectile momentum. 
Our momentum transfer efficiency $b_{\rm eff}$
is not the same as that commonly used to compute the momentum transfer efficiency
resulting from an asteroid impact as 
all the ejecta in our experiments returns to hit the surface, and none of it escapes into space.  Also, the projectile's momentum is a vector but in the right hand side of Equation \ref{eqn:vpert} we integrate the flux of the radial momentum component in the seismic pulse over a hemisphere.  Our dimensionless momentum transfer efficiency includes an integration factor that relates two similar sized quantities with units of momentum.   

Equation \ref{eqn:vpert},  using the maximum projectile deceleration in Equation \ref{eqn:apmax} 
for $dv_p/dt$ and Equation \ref{eqn:alpha_p} for $\alpha_p$, we estimate the peak pulse velocity at $r = R_{cr}$, 
\begin{equation}
v_{pk}(R_{cr}) \sim \frac{b_{\rm eff} C_D}{2} \frac{R_p^2}{R_{cr}^2} \frac{v_{imp}^2}{v_P}. 
  \label{eqn:vpkr_Rc}
\end{equation}
Using our impact velocities, $ b_{\rm eff}C_D=1$, crater sizes listed in Table \ref{tab:exps}, 
the pulse propagation speed listed in Table \ref{tab:pressure} and projectile radii listed in Table \ref{tab:imp}, we estimate that 
 $v_{pk}(R_p) \sim 0.02$ m/s for the millet experiments
and 0.03 m/s for the sand experiment.  These values are similar to those listed in Table \ref{tab:pressure} 
for the accelerometers nearest the impact site and with $r/R_{cr} \sim 1.2$ and 2 for 
the millet and sand experiments, respectively.

We use the peak velocity at $r=R_{cr}$ (Equation \ref{eqn:vpkr_Rc}) to determine the constant of proportionality 
in Equation \ref{eqn:vpkr},
\begin{equation} 
	v_{pk} (r) \approx v_{pk}(R_{cr}) 
	 \left(\frac{r}{R_{cr}} \right)^{-\frac{5}{2}} .
     \label{eqn:vpkr_norm}
\end{equation}

An estimate for 
the peak acceleration at $r=R_{cr}$, given by $a_{pk} \approx v_{pk}/\Delta t_v$,  depends
on the pulse duration at the crater radius or at  $r=R_{cr}$.     We can estimate 
the pulse duration $\Delta t_v$ with Equation \ref{eqn:dtr} or from the time it takes a pressure wave to traverse the crater radius. 
As we discussed in Section \ref{sec:duration}, the diffusion based estimate required a large
$C_W$ scaling coefficient but both estimates gave potential matches to the pulse
duration at $R_{cr}$.  The pulses in our experiments broaden as they
travel, as shown in Figures \ref{fig:broad}.    We opt to use the crater radius
to estimate normalization factors that depend on $\Delta t_v$, giving
\begin{equation}
\Delta t_v(R_c) = \frac{R_{cr}}{v_P},    \label{eqn:dt_assume_0}
\end{equation}
but we adopt
scaling derived from the diffusive model to predict how the amplitudes drop with distance
from the site of impact. 

Using the time it 
takes the pulse to cross the crater radius  for
 the pulse duration $\Delta t_v$ at $r=R_{cr}$ (equation \ref{eqn:dt_assume_0}), we estimate the peak acceleration 
\begin{equation}
 a_{pk}(R_{cr}) \approx \frac{v_{pk}(R_{cr}) v_P}{R_{cr}} .
   \label{eqn:apkr_Rc}
\end{equation}
Using this as the constant of proportionality for Equation \ref{eqn:apkr} the peak acceleration   
\begin{equation}
a_{pk} (r) \approx  a_{pk}(R_{cr})  \left( \frac{r}{R_{cr} } \right)^{-3}.
   \label{eqn:apkr_norm}
\end{equation}
Likewise the peak displacement at $r=R_{cr}$ 
\begin{equation}
\delta_{pk} (R_{cr})  \approx \frac{v_{pk}(R_{cr}) R_{cr}}{v_P}. 
   \label{eqn:dpkr_Rc}
\end{equation}

The exponent of -3  for the decay of the peak acceleration
is nearly consistent with the value of $3.18 \pm 0.10$ measured
for peak accelerations in high velocity impact experiments into quartz sand \citep{Matsue_2020}.

\begin{figure}[htbp] \centering 
\includegraphics[width = 3.3 truein, trim = 20 10 0 0,clip]{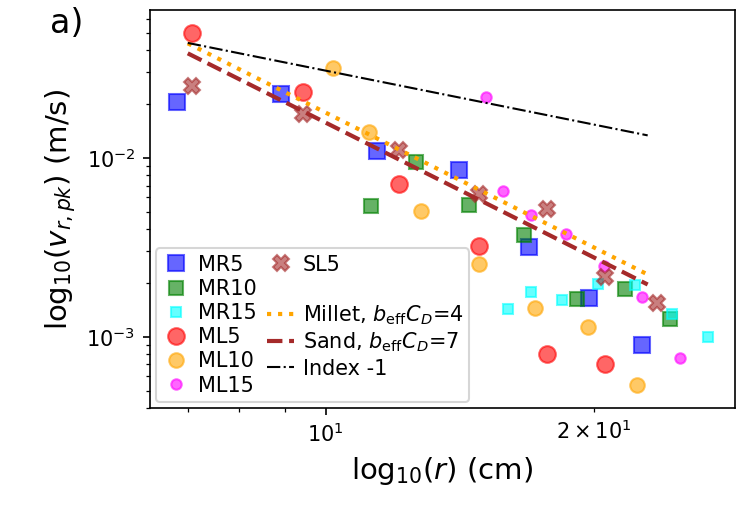}
\includegraphics[width = 3.3 truein, trim = 20 10 0 0,clip]{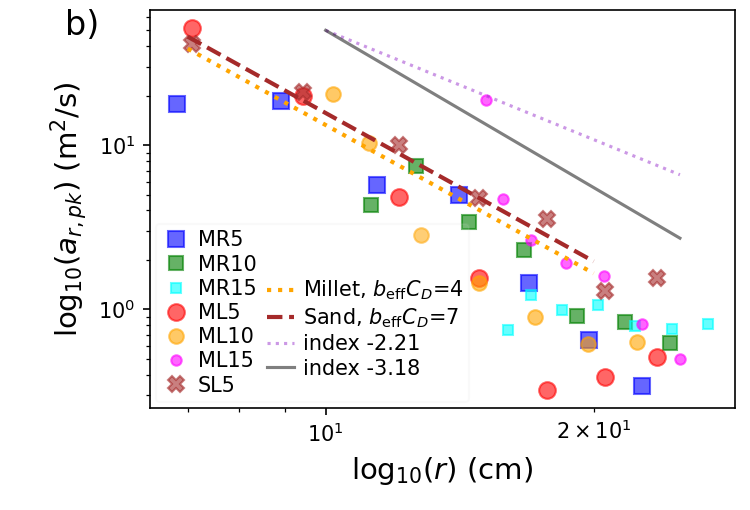}
\caption{
a) Peak pulse velocity as a function of travel distance for the 6 millet and single sand experiments. 
The point types are the same as shown in Figure \ref{fig:speeds}.
Each point is from a single accelerometer. 
The dotted orange line and dashed brown lines are computed
using Equation \ref{eqn:vpkr_norm} with constant of proportionality 
from Equation \ref{eqn:vpkr_Rc}.
The power law index is predicted via Equation 
\ref{eqn:vpkr} using a diffusive attenuation model for pulse propagation. 
The dotted orange line uses parameters for the millet experiments and
the dashed brown line used parameters for sand experiment. 
The product of the momentum transfer efficiency parameter and the drag coefficient $b_{\rm eff} C_D$ has been adjusted to be consistent
with points nearest impact and with values are given in the key. 
The dot-dashed black line shows a model with $v_{pk} \propto r^{-1}$ corresponding to 
little attenuation.  The peak velocities are poorly described by
the black line, so attenuation is rapid.  
b) Peak accelerations as a function of travel distance. 
The dotted orange line and dashed brown lines are similarly computed
using Equation \ref{eqn:apkr_norm} with constant of proportionality 
from Equation \ref{eqn:apkr_Rc}.
The thin solid pink line shows $a_{r,pk} \propto r^{-2.21}$ using the index
measured by \citet{yasui15}.   The thin solid grey line shows  
$a_{r,pk} \propto r^{-3.18}$ measured by \citet{Matsue_2020}.
%
\label{fig:pk}
}
\end{figure}

In Figure \ref{fig:pk} we plot peak velocities and peak accelerations as a function
of distance $r$ from impact site for all accelerometers in the six millet and single
sand impact experiments. The dotted orange and dashed brown lines shows the prediction for
how the peak acceleration and velocity would drop with radius derived
via this simple diffusive attenuation model 
(computed using Equations \ref{eqn:vpkr}, \ref{eqn:apkr}, \ref{eqn:vpkr_Rc}  --  \ref{eqn:apkr_norm}).
The dotted orange lines use impact velocity, projectile and crater radius for the millet experiments
and the dashed brown lines use the same quantities but for the sand experiment.
We have adjusted the produce of the momentum transfer efficiency parameter 
and drag coefficient $b_{\rm eff} C_D$ so
that the lines are consistent with the accelerometers nearest the site of impact.
Our simple diffusion model provides a decent match to the decay rates for the
peak velocities and accelerations in our accelerometers, particularly at positions
nearest the site of impact where reflections don't affect the signal and the signals
are stronger.    
To match the pulse amplitudes we require a momentum transfer efficiency parameter $b_{\rm eff} C_D \sim 4$ or 7, for
the millet and sand experiments, respectively.  We would expect this parameter to be greater than 1 as the seismic pulse must balance the momentum required to both stop the projectile and launch the ejecta curtain and the drag coefficient could exceed 1. 

The power law index of -2.5 predicted for decay of the peak velocity (Equation \ref{eqn:vpkr}) with our simple diffusion model into a granular medium is steeper than the value of -1 measured for the decay exponent for pressure amplitude of sandstone in the elastic regime \citep{Guldermeister_2017}.  An exponent of -1 in the pressure 
amplitude implies that there is little energy lost as the pulse travels. 
With a constant $v_P$, Equation \ref{eqn:P_peak} implies that when there is little attenuation, the peak
velocity $v_{pk}  \propto r^{-1}$ would have a similar exponent.  In Figure \ref{fig:pk} we plot
 a dot-dashed
black line that has $v_{pk}  \propto r^{-1}$.   Pulses in our experiments decay much more rapidly than
shown with this black line.  The -1 exponent for $v_{pk}$ is ruled out by our experiments.

With higher velocity impacts (about 100 m/s) into glass beads \citet{yasui15}
measured an attenuation relation between peak acceleration and distance to impact 
$a_{pk} \propto r^{-2.21  \pm 0.12}$  
(their equation 6).  With impact velocities of 1 to 7 km/s into quartz sand
\citet{Matsue_2020} measured a similar exponent; 
$a_{pk} \propto r^{-3.18  \pm 0.10}$.  
In Figure \ref{fig:pk}b we also plot two lines with these measured indices.  
The decay of peak acceleration we see in our experiments and predicted
with our model is consistent
with the power-law indices measured by these higher velocity impact experiments. 

In the diffusive model, because the pulse broadens, the velocity amplitude decreases even
while conserving momentum (assumed in Equation \ref{eqn:cc}).  This simple model directly relates pulse broadening to energy decay. 
A lossy collisional model gives an analogy.  Each collision redistributes momentum to additional particles, reducing the mean value in a pulse.  
Eventually all particles
have the same velocity.  The energy
decreases, while the distribution of momentum broadens. 

Equations \ref{eqn:vpkr} and \ref{eqn:vpkr_norm} neglect the  dependence of pulse amplitude on polar angle.  We can compare the slope of the points in different experiments in Figure \ref{fig:pk}.  The decay rate of the peak velocity $v_{pk}$ in the ML5 experiment is faster than that of the MR5 experiment.  This difference in decay rates can be attributed to
the angular dependence of the peak velocity.
At the same distance from the site of impact, 
the MR5 experiment has deeper accelerometers than the ML5 experiment
and these deeper accelerometers see higher pulse amplitudes. 

\subsection{Seismic energy decay} 

We estimate how our estimate for the total seismic energy that passes
radius $r$ would scale with radius using the diffusive model discussed above in Section \ref{sec:amp}.  
The seismic energy can be estimated by integrating
the energy flux as a function of time of a pulse that passes radius $r$, giving  
\begin{equation}
E_{seis}(r) \sim \rho_s v_{pk}^2(r) \Delta t_v v_P 2 \pi r^2, \label{eqn:E_seismic2}
\end{equation}
 consistent with Equation \ref{eqn:E_seismic}.
Using Equation \ref{eqn:vpkr} for $v_{pk}$ and Equation \ref{eqn:dtr} for $\Delta t_v$ we find that 
\begin{align}
E_{seis}(r)  & \propto r^{- \frac{5}{2} }  v_P^{-\frac{1}{2}} D^{-\frac{1}{2}} . \label{eqn:EJr}
\end{align}

As done in Section \ref{sec:amp}, we estimate the constant of proportionality
by substituting Equation \ref{eqn:vpkr_Rc} for $v_{pk}(R_{cr})$ and 
with $\Delta t_v (R_{cr}) = R_{cr}/v_P$, 
at crater radius $r=R_{cr}$, 
\begin{align}
E_{seis}(R_{cr}) & \approx \rho_s \left[ \frac{b_{\rm eff} C_D}{2} \frac{R_p^2}{R_{cr}^2} \frac{v_{imp}^2}{v_P}
\right]^2 2 \pi R_{cr}^3 . \label{eqn:EJr_Cr}
\end{align}

\begin{figure}[htbp] \centering 
\if \ispreprint1
\includegraphics[width = 3.5 truein, trim = 20 20 0 0, clip]{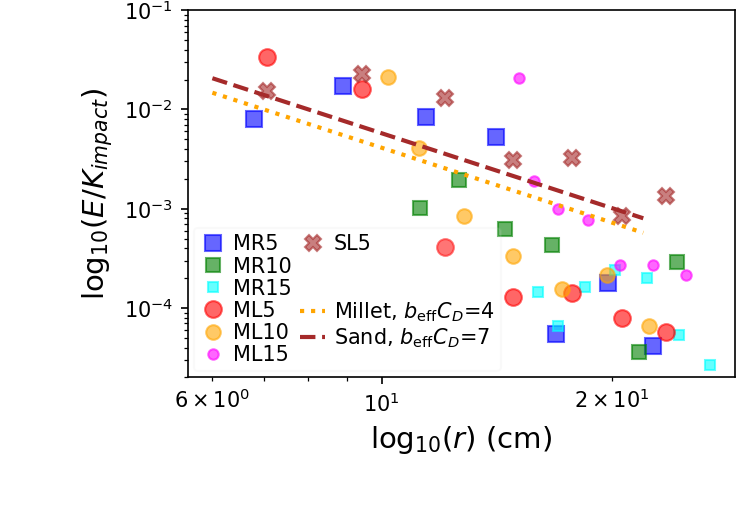}
\else
\includegraphics[width = 4.5 truein, trim = 20 20 0 0, clip]{nov27EJ_new.png}
\fi
\caption{Seismic energy in the pulse as a function
of the distance of each accelerometer from the site of impact.
Total seismic energy is computed for each accelerometer 
using Equation \ref{eqn:E_seismic} and is in units of the kinetic energy of impact. 
The point types are the same as shown in Figure \ref{fig:speeds}.
The dotted orange and dashed lines show the  power law function 
$E_{pk} \propto r^{-\frac{5}{2}}$ 
predicted via Equation \ref{eqn:EJr} and with constant of proportionality from Equation \ref{eqn:EJr_Cr}, computed using parameters for the millet and sand experiments, respectively.
The key shows the assumed momentum transfer efficiency parameters for these two lines. 
\label{fig:EJ}
}
\end{figure}

In Figure \ref{fig:EJ} we show seismic energy estimated from each accelerometer 
using Equation \ref{eqn:E_seismic} 
as a function of distance $r$ from impact site along with a dotted
grey line that shows the $r^{-\frac{5}{2}}$ scaling predicted via Equation \ref{eqn:EJr},
and with constant of proportionality from Equation \ref{eqn:EJr_Cr}.
We use the same momentum transfer efficiencies as in Figures \ref{fig:pk}.
At small radius $r$ from the impact site, the estimated seismic energies exceed the predicted estimate 
because the seismic energies take into a longer time interval than only the first positive portion 
in velocity.   As was true for the peak accelerations and velocities, the estimated decay 
rate is a reasonable match to the measurements from accelerometers nearer the impact site.
Also as was true for peak accelerations and velocities, many points lie below the predicted
line at larger distances.  This discrepancy can in part be attributed to loss of energy through the tub base into the floor. 

\subsection{Extension to higher velocity impacts}
\label{sec:high}

Our power law scaling relations for physical properties have constant of proportionality 
based on physical quantities at the crater radius $R_{cr}$.
However, in section \ref{sec:amp} we based our estimate for the peak velocity at $R_{cr}$ (equation \ref{eqn:vpkr_Rc})
on the product of a drag coefficient, $C_D$, and a momentum transfer parameter $b_{\rm eff}$.
The use of a drag coefficient may only be 
appropriate in a low velocity regime (e.g., \citealt{katsuragi13}).
We are curious to find out if a similar scaling relations are effective in a higher velocity impact regime. 
We explore two approaches for estimating physical quantities at the crater radius,
first, using the projectile's kinetic energy $K_{imp}$ and a seismic efficiency parameter $k_{seis}$, 
and second, using the projectile's momentum  $m_p v_{imp}$ and a different momentum transfer parameter 
$B_{\rm eff}$ to characterize the fraction
of projectile momentum that is transferred into the seismic pulse.
With these two approaches we estimate the peak velocity  $v_{pk}(R_{cr})$
and  peak acceleration $a_{pk}(R_{cr})$ at the 
crater radius.  We use  $a_{pk}(R_{cr})$  to normalize the peak accelerations
we measure in our experiments and to compare our experiments
to the higher impact velocity experiments by  \citet{yasui15} and \citet{Matsue_2020}, 
who also measured peak accelerations, which they called $g_{max}$. 

Because a seismic source time gave pulse duration similar to that observed
in both our experiments and the higher impact velocity experiments
 by \citet{yasui15} and \citet{Matsue_2020}, we can use it to estimate
the pulse duration at a distance of the crater radius 
as in equation \ref{eqn:dt_assume_0}.  

Using Equation \ref{eqn:E_seismic2} for the seismic energy
at a distance $r=R_{cr}$ and with a pulse width from equation \ref{eqn:dt_assume_0}, the seismic energy  
\begin{align}
E_{seis}(R_{cr} ) & = k_{seis} K_{imp} \label{eqn:E_DART_0} \\
& \sim \rho_s v_{pk}^2(R_{cr})  2 \pi R_{cr}^3. \label{eqn:E_DART1}
\end{align}
This gives us a peak pulse velocity at the crater radius 
\begin{align}
v_{pk}(R_{cr}) \sim &  \sqrt{ \frac{ k_{seis} K_{imp}}{ 2 \pi R_{cr}^3 \rho_s}} . \label{eqn:vpk_new1}
\end{align}
Using equation \ref{eqn:dt_assume} and $a_{pk} \sim v_{pk}/\Delta t_v$, 
equation \ref{eqn:vpk_new1} gives peak acceleration 
\begin{align}
a_{pk}(R_{cr}) \sim &  \sqrt{ \frac{ k_{seis} K_{imp}}{ 2 \pi R_{cr}^3 \rho_s}} \frac{v_P}{R_{cr}}.
\label{eqn:apk_new1}
\end{align}
The seismic efficiency parameter $k_{seis}$ is the same as other studies have used to characterize
the strength of a seismic pulse (e.g., \citealt{mcgarr69,shishkin07,yasui15,Guldermeister_2017,Matsue_2020}), however, because the energy in the pulse decays as it propagates, 
it is dependent upon the distance from impact site at which it is calculated.   
Here $k_{seis}$ refers to the value
at $r = R_{cr}$, the crater radius.

Instead of scaling from seismic energy, we could use the projectile's  momentum $m_p v_{imp}$.
Following Equation \ref{eqn:cc} for the momentum in the pulse  
\begin{align}
B_{\rm eff} m_p v_{impact} \sim \rho_s v_{pk}(R_{cr})  \Delta t_v (R_{cr})v_P(R_{cr})
 2 \pi R_{cr}^2 . 
  \end{align} 
This with Equation \ref{eqn:dt_assume_0} for the pulse duration at $R_{cr}$ gives 
\begin{align}
v_{pk}(R_{cr}) & \sim  \frac{ B_{\rm eff} m_p v_{imp} }{ 2 \pi R_{cr}^3 \rho_s } \label{eqn:vpk_new2}
 \end{align} 
 and 
 \begin{align}
a_{pk}(R_{cr}) & \sim  \frac{ B_{\rm eff} m_p v_{imp} }{ 2 \pi R_{cr}^3 \rho_s }  \frac{v_P}{R_{cr}}.
\label{eqn:apk_new2}
 \end{align} 
The momentum transfer parameter $B_{\rm eff}$ is similar to but not identical to the  $\beta$
parameter used to characterize the ratio of momentum transferred to an asteroid following an impact
to the projectile momentum 
\citep{housen11,Holsapple_2012,Jutzi_2014,Flynn_2015}.  
Our parameter $B_{\rm eff}$ characterizes the ratio of the radial component of momentum in the seismic pulse to that in the projectile at $r= R_{cr}$.
 
\begin{figure}\centering 
\if \ispreprint1
\includegraphics[width=3.5 truein,trim = 10 20 0 0, clip]{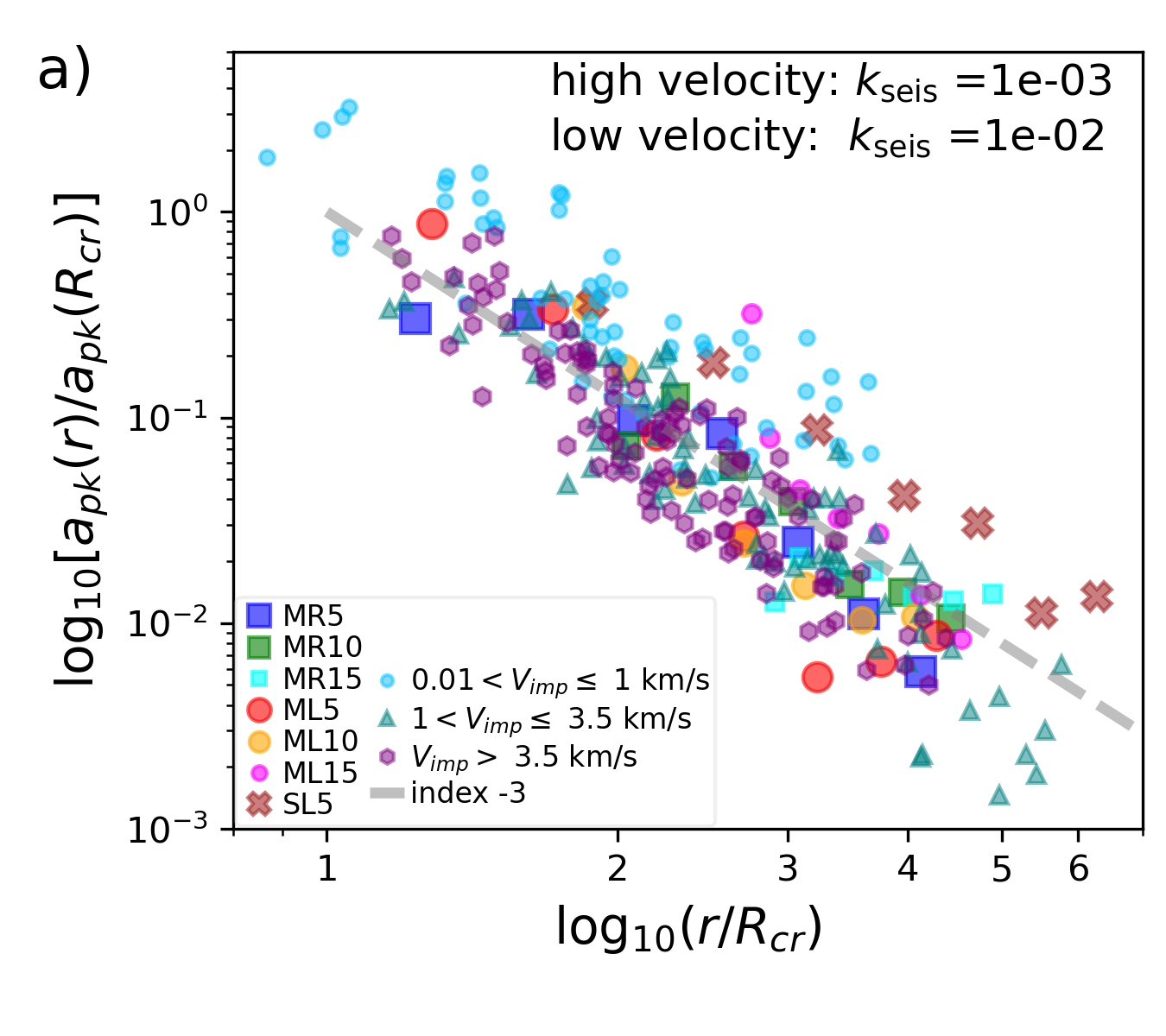}
\includegraphics[width=3.5 truein,trim = 10 20 0 0, clip]{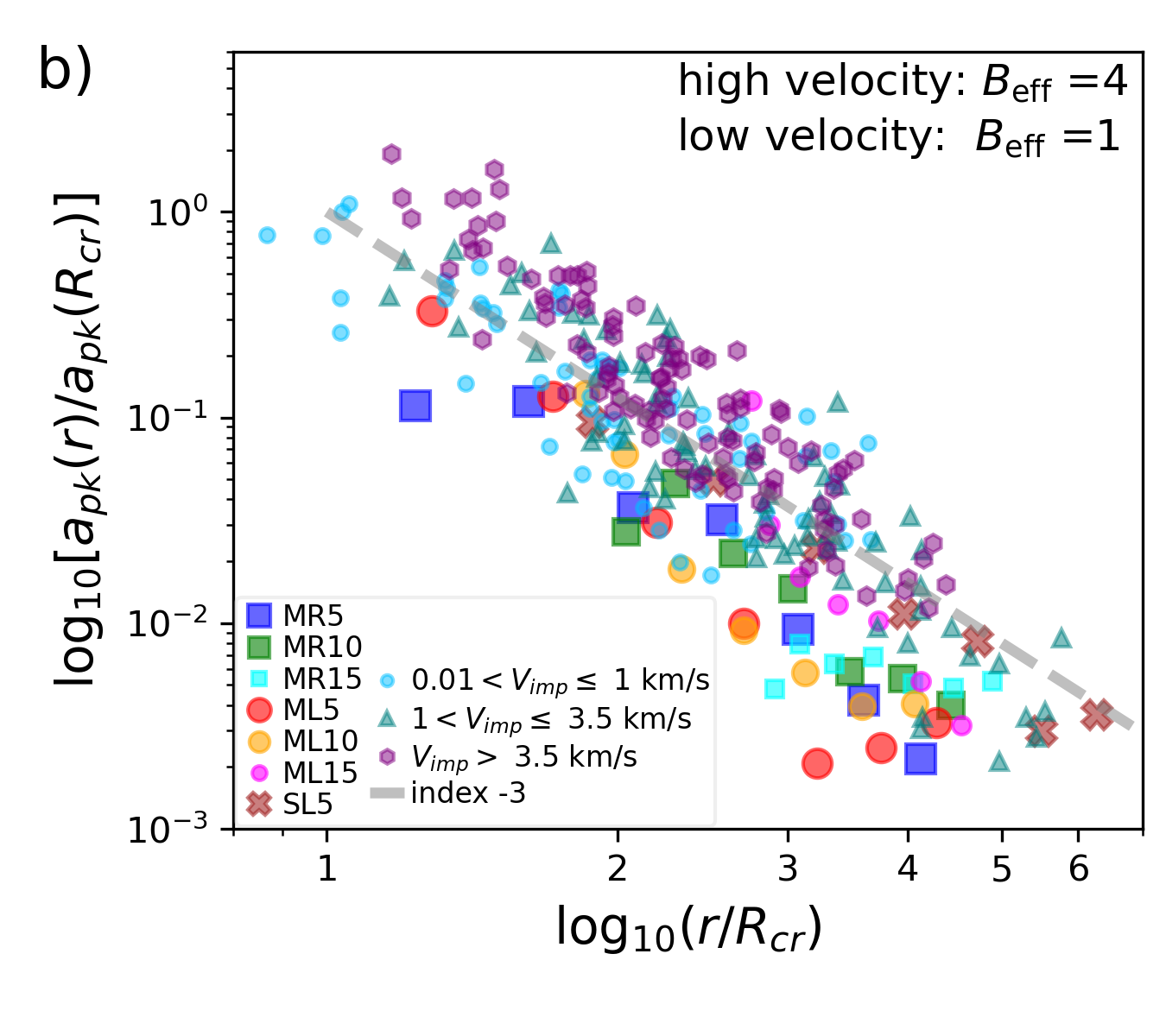}
\else
\includegraphics[width=3.0 truein,trim = 10 20 0 0, clip]{YM_energy.png}
\includegraphics[width=3.0 truein,trim = 10 20 0 0, clip]{YM_momentum.png}
\fi
\caption{a) Normalized peak accelerations as a function of distance from impact divided by crater radius. 
The peak accelerations have been divided by $a_{pk}(R_{cr})$ computed using equation 
\ref{eqn:apk_new1} and with impact kinetic energy, crater radius, pulse travel velocity and substrate density for each experiment.  The seismic efficiency parameter $k_{seis}$ we use for our low velocity experimental measurements  
and for the high velocity experimental measurements by  \citet{yasui15} and \citet{Matsue_2020}
are printed on the upper right and are also listed in Table \ref{tab:cal}. 
The dashed grey line shows $a_{pk}(r)/a_{pk}(R_{cr}) = (r/R_{cr})^{-3}$.
The point types for our data (the low velocity experiments) are the same as shown in Figure \ref{fig:speeds}.
The point types for the high velocity experiments are the same as in Figure \ref{fig:TT}.
b) similar to a) except equation \ref{eqn:apk_new2}  is used to compute the normalization 
for the acceleration.  This depends on the momentum transfer parameter $B_{\rm eff}$ which 
is printed on the upper right. 
\label{fig:YM}}
\end{figure}

In Figure \ref{fig:YM}a we plot peak accelerations from our experiments as a function 
of distance from impact in units of the crater radius $R_{cr}$.
On this plot we include experimental measurements for the peak accelerations
from the higher velocity impact 
experiments by \citet{yasui15} and \citet{Matsue_2020}.  
The peak accelerations are normalized by $a_{pk}(R_{cr})$
from  equation \ref{eqn:apk_new1} which is computed 
with the impact kinetic energy, crater radius, pulse travel velocity and substrate density appropriate for each experiment.     
The point types for our data are the same as shown in Figure \ref{fig:speeds}.
The dashed grey line
shows  $a_{pk}(r)/ a_{pk}(R_{cr})=(r/R_{cr})^{-3}$ with power law index predicted with equation 
\ref{eqn:apkr_norm}. 
We have adjusted the dimensionless seismic efficiency $k_{seis}$ so that the data
lie near this dashed line.  We used a seismic efficiency of $k_{seis}=10^{-2}$ for our low
velocity experimental measurements
and $k_{seis}=10^{-3}$ for the higher velocity experiments. The higher seismic efficiency we estimate 
in our experiments is consistent with our previous discussion in section \ref{sec:seismic}.

Figure \ref{fig:YM}b is similar to Figure \ref{fig:YM}a except equation \ref{eqn:apk_new2}  is used
to normalize the accelerations.   For this figure we adjusted the momentum transfer parameter
$B_{\rm eff}$ so that the data points lie near the grey dashed line.  
Figure \ref{fig:YM} illustrates that our power law index of -3, predicted via the
advection diffusion model, is within an order of magnitude consistent with accelerations 
experimentaly measured in  
both the high and low velocity impact velocity regimes. Within 
an order of magnitude, we find that we can estimate the amplitude of the impact generated pulse using 
either the projectile momentum or its kinetic energy in both low and high impact velocity regimes. 

In section \ref{sec:amp} we based our estimate for the peak velocity at $R_{cr}$ (equation \ref{eqn:vpkr_Rc}) on a drag coefficient, $C_D$ and a momentum transfer
parameter $b_{\rm eff}$.  For our low velocity experiments, we estimated
that properties at the crater radius are consistent with $b_{\rm eff} C_D \sim 4$ or 7, depending
upon whether the experiments were in millet or sand.
If estimate the peak accelerations 
at the crater radius using
equations \ref{eqn:vpkr_Rc} and \ref{eqn:apkr_Rc},
we find that the product of the momentum transfer parameter 
and drag coefficient $b_{\rm eff} C_D$  must be strongly dependent upon impact velocity.
By creating
a figure similar to Figure \ref{fig:YM} but using equation \ref{eqn:apkr_Rc}, we find that 
for impact velocities of $\sim 100$ m/s we estimate $b_{\rm eff} C_D \sim 0.1$ using the
data by \citet{yasui15}. 
The high velocity impact velocity measurements by \citet{Matsue_2020} would require 
$b_{\rm eff} C_D \sim 0.005$.  
We find that equations \ref{eqn:vpk_new1} and \ref{eqn:vpk_new2}
are superior to equation \ref{eqn:vpkr_Rc} in the sense that they are approximately correct
over a larger range of impact velocity. 

Our estimates for the dimensionless parameters $b_{\rm eff} C_D$, $B_{\rm eff}$ and
$k_{seis}$ for the experiments 
at different impact velocities are summarized in Table \ref{tab:cal}.
Here slow impacts are those with impact velocity lower than 10 m/s
and the parameter estimates are based on our experiments. 
Fast impacts are those with 100 m/s to 5 km/s impact velocities  and based on
 the experiments
by \citet{yasui15} and \citet{Matsue_2020}.

\begin{table}[htbp]  \centering
\caption{Values for estimating quantities at the crater radius \label{tab:cal}}
\begin{tabular}{lllll}
\hline
Parameter  & $b_{\rm eff} C_D$ & $k_{seis}$ & $B_{\rm eff}$ \\
\hline
Description   
	& \begin{tabular}{@{}l@{}}Momentum \\ transf. times \\ drag coeff.\end{tabular}  
	& \begin{tabular}{@{}l@{}}Momentum \\ transfer \\ efficiency \end{tabular} 
	&  \begin{tabular}{@{}l@{}}Seismic \\ efficiency\\ parameter \end{tabular}  \\
Expression   & Eqn. (\ref{eqn:vpkr_Rc}) & Eqn. (\ref{eqn:vpk_new1}) & Eqn. (\ref{eqn:vpk_new2}) \\
Slow impacts & 4 to 7 & $10^{-2}$ & 1 \\
Fast impacts & 0.1 to 0.005 & $10^{-3}$ & 4 \\
\hline
\end{tabular}
\end{table}

\section{The pulse excited by the DART impact on Dimorphos}
\label{sec:DART}

In Sept. 2022, the DART spacecraft will impact the 
secondary of the asteroid (65803) Didymos system, known as Dimorphos.
Using the diffusive model developed in Section \ref{sec:diff}, we estimate the characteristics
of a seismic pulse that is driven by the impact. 
The DART impact velocity will be about $v_{imp} \approx 6.5$ km/s and the impactor 
mass at the time of impact is
expected to be about $m_p \approx 550$ kg \citep{Rivkin_2021}.  This gives 
a kinetic energy of $K_{imp} \approx 1.2 \times 10^{10} $J.

The mass and radius of the target, Dimorphos,
 are estimated to be $m_a \approx 5 \times 10^9$ kg and  
$R_a \sim 80 $ m \citep{Naidu_2020}.  
Dimorphos' density is based on estimates of the density of the primary, Didymos,   
$\rho_s \approx 2.17$ g cm$^{-3}$ \citep{Naidu_2020}. 
The gravitational acceleration on the surface
of Dimorphos $g_{\rm eff} \sim Gm_a/R_a^2 \sim  5 \times 10^{-5} {\rm m \ s}^{-2}$.

The material properties of Dimorphos could resemble those of the recently well characterized
rubble NEA asteroid (101955) Bennu.
The bulk mass density of Bennu is $\rho = 1.190 \pm$ 13 g cm$^{-3}$
\citep{Lauretta_2019}. 
\citet{Tricarico_2021} estimate a 
rubble bulk density (that of the rubble pieces inside the asteroid) 
of 1.35 g/cm$^3$.  This is somewhat higher than the bulk density of
the asteroid itself 1.19 g/cm$^3$ including voids, so \citet{Tricarico_2021} 
estimate moderate macro-porosity of about 12\% (corresponding to voids between rocks).
As the rubble bulk density
is lower than that of representative meteoritic samples, $\sim 3$ g/cm$^3$ 
(see \citealt{Macke_2011}),
\citet{Tricarico_2021}  infer that the rocks themselves are likely to exhibit micro-porosity $\sim $ 50\%.
This high porosity is similar to the porosity of some meteoritic samples \citep{Macke_2011}.
Seismic wave velocity for rocks is reduced, compared to 
that of solid constituents, due to porosity and the nature and density of internal voids and cracks \citep{OConnell_1974,Kovacik_1999}.    
Because Dimorphos's rocks are likely to be porous, we estimate the elastic modulus of the grain material to be low,  
$E_{g} \sim 10$ GPa, and similar to that of some sandstones. 
At a density of 2 g/cm$^{-3}$, this gives a sound speed within each rock or grain of 
$c_g \sim \sqrt{E_g/\rho_s} \sim 2000 $ m/s.

Because Dimorphos is a low-g environment, 
pulse propagation should be a regime with velocity that depends on peak pressure in 
the pulse, so we adopt the pressure dependent scaling by \citet{vandenWildenberg_2013} which is 
based on experiments into polydisperse glass beads.
The relation is 
\begin{equation}
\frac{v_P}{c_g} \sim  \left( \frac{P_{pk}}{E_g}  \right)^\frac{1}{6}. \label{eqn:1_6}
\end{equation}
To order of magnitude this relation is similar to the $v_P/c_g$ vs $P_0/E_g$ relation 
by \citet{Jia_1999} for disordered glass spheres that depends
upon confinement pressure rather than peak pressure.
Equation \ref{eqn:1_6} along with Equation \ref{eqn:P_peak}, 
gives a relation between peak pressure and peak velocity in a traveling longitudinal pressure pulse 
\begin{equation}
P_{pk} \sim \left[ \rho_s v_{pk} c_g \right]^\frac{6}{5} E_g^{- \frac{1}{5}}.
\label{eqn:P_peak2}
\end{equation}

With gravity inversion methods, \citet{Tricarico_2021} find that 
 the rubble size frequency distribution of the interior of asteroid (101955) Bennu
is consistent with that observed on the surface, with a 
cumulative index of approximately $-2.9$ \citep{DellaGiustina_2019}.
With this index, the mean particle size is set by the smallest particles
whereas the mean particle volume is set by the largest particles.
Since, the largest particles carry the strong force chains \citep{Voivret_2009},
the larger particles could set the properties of seismic attenuation within the medium.
The particle size distribution seen on the surface has a number of particles
 with diameters of  50 m, so we use that as a size-scale to estimate a 
diffusion coefficient related to pulse broadening. 
For a fiducial value, we use a diffusion coefficient in the form
of Equation \ref{eqn:D} with $D/v_P = C_W^2 d_g = 10^3 $ m.  We  use a large
value so as to be consistent with the values we estimated for $C_W \sim 10$ from
our experiments described in sections \ref{sec:duration}
and \ref{sec:diff}. 

Localization of seismic waves generated by hypervelocity (velocity of greater than a km/s) impacts into sandstone gives impact site as origin \citep{Moser_2013}.  Even
though our experiments show anisotropy in pulse strength, we adopt a model where
pulse strength only depends on travel distance $r$ from the site of impact (and following \citealt{thomas05}). 
We assume that 
the transition to a propagating seismic pulse takes place at a distance near the crater radius.
 As done in Section \ref{sec:diff}, we scale properties of the travel seismic pulse from the crater radius, 
 at a distance $r = R_{cr}$. 
The radius of the impact crater on Dimorphos that will be caused by the DART impact 
 has been estimated with scaling laws \citep{housen11} and ranges from 
$R_{cr} \sim $ 4 to 45 m, with value that 
depend upon the substrate material properties \citep{Cheng_2016,Cheng_2020}.
For a fiducial value we adopt $R_{cr} = 10$ m.

Because a seismic source time gave pulse duration similar to that observed
in both our experiments and those by \citet{yasui15} and \citet{Matsue_2020}, we use it to estimate
the pulse duration at a distance of the crater radius from the site of the DART impact. 
We assume that the pulse width at the crater radius is  
\begin{equation}
\Delta t_v(R_{cr}) \sim  \frac{R_{cr}}{v_P(R_{cr}) }.  \label{eqn:dt_assume}
\end{equation}  

The seismic efficiency in the high velocity DART impact should be lower than the seismic efficiency
measured in our low velocity experiments because energy would be lost during shock propagation and
while the pressure wave propagates in a plastic regime \citep{shishkin07,Guldermeister_2017}.
We adopt a fiducial value for the seismic efficiency $k_{seis} \sim 10^{-3}$, 
following the estimates and measurements by \citet{shishkin07,Guldermeister_2017}
and approximately consistent with the high velocity laboratory impacts 
by \citet{yasui15} and \citet{Matsue_2020}. 
%
Equation \ref{eqn:E_seismic2} for the seismic energy
at a distance $r=R_{cr}$ and  pulse width from equation \ref{eqn:dt_assume},  gives 
equations \ref{eqn:E_DART_0} and \ref{eqn:E_DART1},   and these give 
peak pulse velocity at the crater radius in equation \ref{eqn:apk_new1}.
Using values for the DART impact and equation \ref{eqn:apk_new1}
we estimate the peak pulse velocity at the crater radius 
\begin{align}
v_{pk}(R_{cr}) \sim &  \sqrt{ \frac{ k_{seis} K_{imp}}{ 2 \pi R_{cr}^3 \rho_s}} \nonumber \\
 = &1.0 \ {\rm m/s} 
 \left( \frac{k_{seis}}{10^{-3} } \right)^\frac{1}{2}  \! 
  \left( \frac{K_{imp}}{1.2 \times 10^{10} {\rm J} } \right)^\frac{1}{2}  \times \nonumber \\
& \ \   \left( \frac{\rho_s}{2000 \ {\rm kg\ m}^3} \right)^{-\frac{1}{2} } \!
  \left( \frac{R_{cr}}{10 \ {\rm m} } \right)^{-\frac{3}{2}} .  \label{eqn:vpk_DART}
\end{align}

Instead of scaling from seismic energy, we could use the momentum of impact $m_p v_{imp}$
and a momentum transfer efficiency parameter $B_{\rm eff}$ to characterize the fraction
of momentum that is transferred into a seismic pulse.   
Hypervelocity impact experiments into pumice estimate a recoil from crater ejecta that
exceeds the direct momentum transferred by absorption of the projectile by a factor of order unity 
(e.g.,  \citealt{Flynn_2015}), though $B_{\rm eff} \sim 4 $  is more consistent with
the high velocity laboratory experiments by \citet{yasui15} and \citet{Matsue_2020}, 
as we discussed in section \ref{sec:high}.
Equation \ref{eqn:cc} for the momentum in the pulse and using equation \ref{eqn:dt_assume} 
for the pulse width, gives 
a peak velocity at the crater radius in equation \ref{eqn:apk_new2}.  
Using values for the DART impact 
\begin{align}
v_{pk}(R_{cr}) & \sim  \frac{ B_{\rm eff} m_p v_{imp} }{ \rho_s 2 \pi R_{cr}^3 } \nonumber \\
 & \sim 0.28 {\rm m/s} 
 \left( \frac{B_{\rm eff}}{1 } \right) \! 
  \left( \frac{m_p v_{imp} }{3.6 \times 10^{6}\ {\rm kg\ m \ s}^{-1} } \right) \!  \nonumber \\
 &  \ \  \left( \frac{\rho_s}{2000 \ {\rm kg\ m}^3} \right)^{-1} \!\!
    \ \  \left( \frac{R_{cr}}{10 \ {\rm m}} \right)^{-3} .
\end{align} 
As this is similar in size to Equation \ref{eqn:vpk_DART}, either the 
momentum transfer efficiency parameter $B_{\rm eff}$ or seismic efficiency $k_{seis}$ can be used to estimate the peak velocity
$v_{pk}(R_{cr})$. 
Equations \ref{eqn:1_6}, \ref{eqn:P_peak2}, \ref{eqn:dt_assume} and \ref{eqn:vpk_DART} are sufficient to estimate other physical quantities at $r=R_{cr}$. 
Pulse amplitude and other quantities at the crater radius approximately scale with our initial choice of 
crater radius in the following manner; 
$v_{pk}(R_{cr}) \propto R_{cr}^{-\frac{3}{2}}$ (from equation \ref{eqn:vpk_DART}), 
$P_{pk}(R_{cr}) \propto R_{cr}^{-\frac{9}{5}}$ (following equation \ref{eqn:P_peak2}), 
$v_P(R_{cr}) \propto R_{cr}^{-\frac{3}{10}}$ (following equation \ref{eqn:1_6}),
$\Delta t_v (R_{cr})  \propto R_{cr}^\frac{13}{10}$ (following equation \ref{eqn:dt_assume}), 
$a_{pk}(R_{cr}) \sim v_{pk}/\Delta t_v \propto R_{cr}^{-\frac{14}{5}} $, and 
$\delta_{pk}  (R_{cr}) \sim v_{pk} \Delta t_v \propto R_{cr}^{-\frac{1}{5}}$. 
The seismic energy
is fixed by the seismic efficiency  (following equation \ref{eqn:E_DART_0}).

If attenuation is directly related to pulse duration, then with an assumption for the form of the diffusion coefficient,
we can estimate how the generated pulse broadens as it travels.
Using the diffusive model discussed in Section 
\ref{sec:diff},  the square of the pulse duration for a pulse that propagates from $r = R_{cr}$ 
\begin{align}
[\Delta t_v (r)]^2 & = [\Delta t_v (R_{cr})]^2 + \int_{R_{cr}}^r dr \frac{D}{v_P(r)^3} .
\label{eqn:int_dt}
\end{align}
We use diffusion coefficient $D$  in the form of Equation \ref{eqn:D}
 for a propagation velocity that depends on pulse amplitude (or the peak pressure in the pulse).  

\begin{figure}[htbp] \centering 
\if \ispreprint1
\includegraphics[width=3.3 truein, trim = 20 20 0 0, clip]{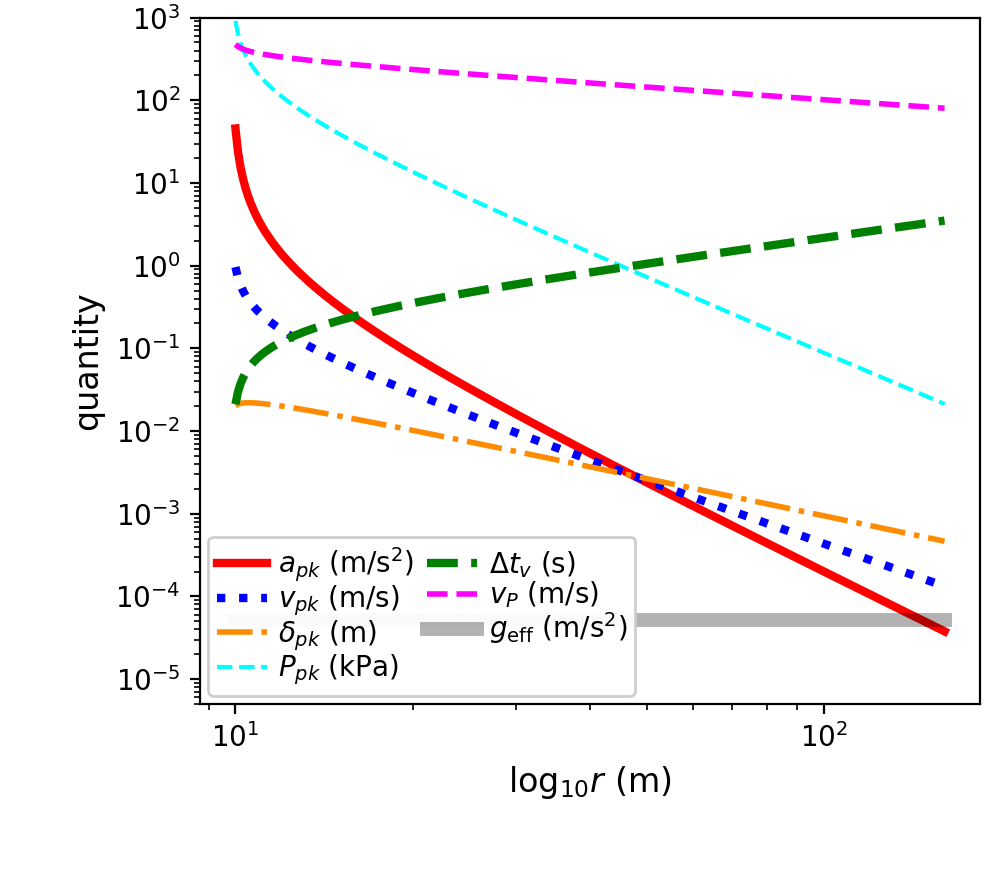}
\else
\includegraphics[width=4.3 truein, trim = 20 20 0 0, clip]{DART_quants.png}
\fi
\caption{
Quantities estimated for a pulse excited by the DART impact into Dimorphos 
as a function of distance from the site of impact. 
Parameters for this model are listed in Table \ref{tab:DART}.
The model was created by integrating Equation  \ref{eqn:int_dt} and using Equations  \ref{eqn:Ppkr_DART}, \ref{eqn:1_6}, \ref{eqn:P_peak2} and values at the crater radius based on Equations  \ref{eqn:dt_assume} and  \ref{eqn:vpk_DART}.
The thick grey line shows the surface gravity of Dimorphos, $g_{\rm eff}$ in m/s$^2$.
 \label{fig:DART}
 }
\end{figure}

As in Section \ref{sec:diff}, we assume that the radial component
of momentum per unit solid angle is approximately conserved while the pulse propagates.
Equation \ref{eqn:cc} and Equation \ref{eqn:P_peak} give peak pulse pressure 
\begin{align}
P_{pk}(r) & \sim P_{pk}(R_{cr}) 
\left(\frac{R_{cr}}{r} \right)^{2} 
\left(\frac{\Delta t_v(R_{cr})}{\Delta t_v(r)} \right). 
\label{eqn:Ppkr_DART}
\end{align} 
By integrating Equation \ref{eqn:int_dt} while updating the peak pressure $P_{pk}$ with Equation \ref{eqn:Ppkr_DART}, and using Equations \ref{eqn:1_6} and \ref{eqn:P_peak2} for $v_P$ and $P_{pk}$, 
we can compute physical quantities such as $v_P$, $v_{pk}$ as a function of  distance from the site of impact. 
A resulting integration is shown in Figure \ref{fig:DART} with fiducial parameters summarized in
Table \ref{tab:DART}.
In this figure we plot pulse peak pressure $P_{pk}$, 
peak acceleration $a_{pk}$, peak velocity $v_{pk}$, peak displacement $\delta_{pk}$,
pulse velocity $v_P$ and pulse duration $\Delta t_v$.  These quantities are plotted as a function
of travel distance from the site of impact $r$.  We only plot quantities for $r> R_{cr}$ as 
pulse properties are scaled from our estimates at the crater radius, $r = R_{cr}$.
The plot is limited to  $ r < 2 R_a$ where $R_a$ is the estimated radius of Dimorphos.  

The fiducial model in Figure \ref{fig:DART} 
shows that the diffusive spreading rapidly dominates the pulse duration.
Past about $r \sim 2 R_{cr}$,  
the pulse width approximately scales with $r^\frac{1}{2}$,  
similar to that obeyed by our model in Section \ref{sec:diff} where pulse velocity $v_P$
is constant.  Here the pulse travel speed decreases as the pulse travels,  but because it is only
weakly dependent on pulse pressure (to the 1/6--th power), $v_P$ only varies slowly with $r$. This implies that the power law indices for pulse properties that we previously estimated in Section \ref{sec:diff} as a function of travel distance $r$ will also be obeyed in the model discussed here; $v_{pk}, P_{pk}, E_{seis} \propto r^{- \frac{5}{2}}$, $a_{pk} \propto r^{-3}$, $\delta_{pk} \propto r^{-2}$, and $\Delta t_v \propto r^{\frac{1}{2}}$.  This also implies that the magnitudes of the estimated physical quantities are
not strongly dependent upon the assumed index of 1/6. 

From Figure \ref{fig:DART}, we estimate
a surface displacement in the seismic pulse of only a few cm at $r \approx R_{cr}$.
LICIACube will have a flyby distance of 55 km, and its imagery should have a resolution at closest
approach of 1.4 m pixel$^{-1}$ \citep{Dotto_2021}.   
Boulder displacements in the seismic pulse itself are a few cm large at $r=R_{cr}$ and so would be about two orders of magnitude out of reach of LICIACube's imaging. 
Accelerations caused by the seismic pulse should be significant compared to the surface gravity.
For our fiducial model, $a_{pk} > g_{\rm eff}$, peak acceleration exceeds the surface gravity of the asteroid (about $0.5 \times 10^{-4}$ m/s$^2$), for a travel length equal to asteroid diameter.
Despite rapid attenuation, particles on the surface would be perturbed all over the asteroid.
Only near the crater would peak pulse velocities be similar to the escape velocity, which on Dimorphos is about 10 cm/s, as expected for formation of the crater. 

\begin{figure}[htbp]  \centering
\if \ispreprint1
\includegraphics[width=3.3 truein, trim = 20 20 0 0, clip]{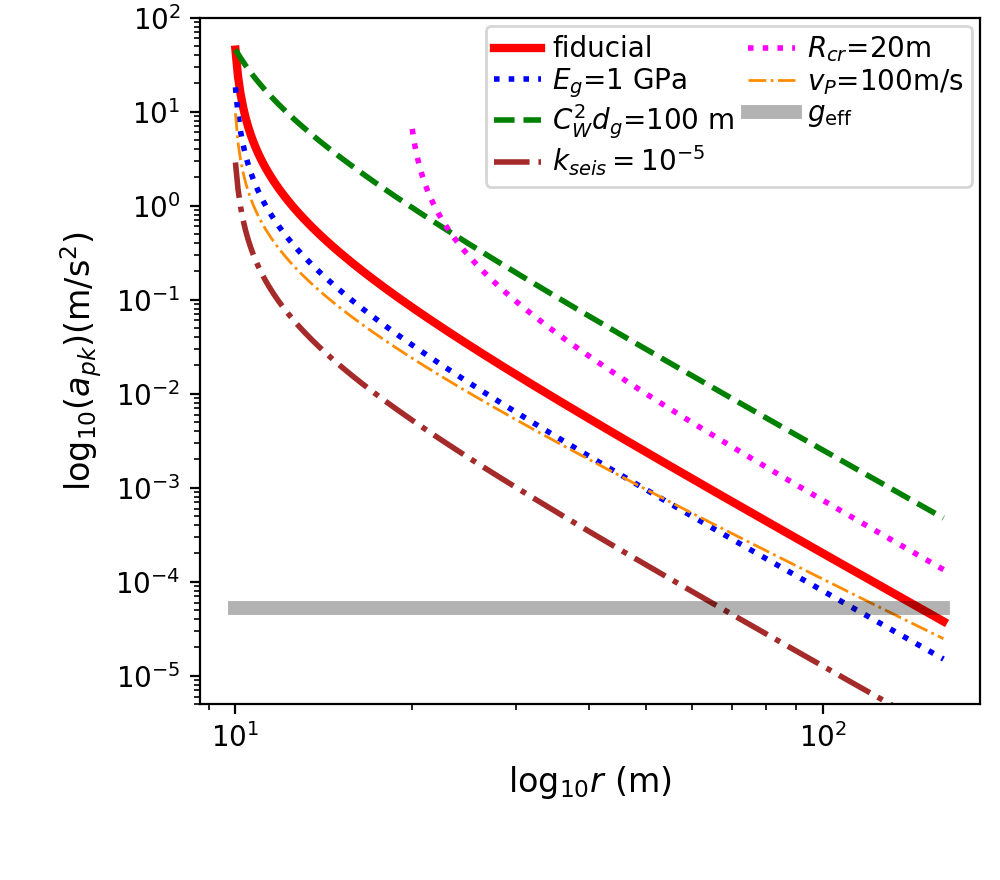}
\else
\includegraphics[width=4 truein, trim = 20 20 0 0, clip]{DART_ac.png}
\fi
\caption{
Pulse peak acceleration as a function of distance from impact for DART impulse models.  
The fiducial model, here shown with a solid red line, is the same as that shown in Figure \ref{fig:DART} and with parameters
listed in Table \ref{tab:DART}.  
The other lines show models that are like the fiducial model except 
a single parameter is changed.  The new parameter value for each model is shown in the key.
We also show a diffusive and constant pulse velocity ($v_P$) model.
The thick grey line shows the gravitational acceleration at the surface of Dimorphos. 
 \label{fig:DART_ac}
 }
\end{figure}

In Figure \ref{fig:DART_ac} we show peak accelerations for models similar to the fiducial one.
We vary a single parameter in each model.   We also show a constant pulse velocity model with $v_P= 100$ m/s that is independent of pulse amplitude.
Reducing the grain elastic modulus has the same effect as reducing $v_P$.  
The pulse is stronger with a smaller diffusion coefficient and if the crater radius is large.  
All of the models shown in Figure \ref{fig:DART_ac} show peak accelerations above the surface
effective gravity $g_{\rm eff}$ for travel distances that are similar to the radius of the asteroid.
Despite the high level of attenuation implied by our diffusive model,  the seismic pulse
traverses the asteroid and can cause significant surface motions that can disturb the surface.  

When they reach a granular surface, a strong seismic pulse can cause the surface to deform, induce landslides and loft particles off the surface (e.g., \citealt{tancredi12,wright19}).  These processes would also reduce the amplitude of reflected waves and increase the attenuation rate.  After the pulse has traversed the asteroid, seismic energy would continue to attenuate and peak accelerations would drop below $g_{\rm eff}$.  The impact generated pulse would only be strong enough to disturb the surface during a single crossing time.  
The scenario is consistent with the seismic jolt model used to account for crater erasure
on Eros with a single seismic pulse \citep{thomas05}. 

\begin{table}[htbp]  \centering
\caption{DART impulse fiducial model}
\label{tab:DART}
\begin{tabular}{lllllll}
\hline
Quantity &   & Value \\
\hline
Kinetic energy of impact & $K_{imp}$ & $1.2 \times 10^{10}$ J \\
Seismic efficiency & $k_{seis}$ & $10^{-3}$ \\
Bulk density & $\rho_s$  & 2000 kg m${^{-3}}$ \\
Substrate elastic modulus  & $E_g$  & 10 GPa \\
Crater radius & $R_{cr}$ & 10 m \\
Diffusion coef. parameter & $\frac{D}{v_P}  = C_W^2 d_g$\!\!\!\!& 1000 m \\
\hline
\end{tabular}
\end{table}

\section{Summary and Discussion}
\label{sec:sum}

In this paper we have used an array of 7  accelerometers to characterize propagation and decay 
of a pulse excited in a granular medium by a low velocity normal impact.   Our impact velocity
is about 5 m/s and our impacts take place into millet or fine sand that fill an approximately 
cylindrical 41.6 litre tub.  Our experiments are complimentary to those by \citet{yasui15} and 
\cite{Matsue_2020} as they are at lower velocity, in different granular substrates, and measure 
pulse properties at more accelerometer positions.  

The accelerometers primarily detect a longitudinal pressure pulse that propagates radially
away from the site of impact.  We estimate that a seismic pulse
is launched within a few ms from the moment of impact.   
The pulse travel speed is similar in millet and fine sand, and about 55 m/s and matches
that seen by \citet{Matsue_2020} in 100 m/s impact experiments into quartz sand. 
The pulse travel speed in units of the sound speed within the grains for both substrates is approximately 
consistent with predictions and experimental measurements of pressure dependent wave
propagation in a disordered granular medium 
\citep{Jia_1999,Somfai_2005,vandenWildenberg_2013}.

We examined the orientation of the peak accelerations.  Acceleration vectors
are oriented radially with origin at the site of impact. 
This implies that we primarily see longitudinal wave propagation.  
If asteroid material behaves similarly, 
we would not expect to see phenomena
on rubble asteroids that is associated with shear waves, such as Rayleigh waves or antipodal focusing, though jamming associated with shear stress \citep{Bi_2011} could be relevant for pulse generation in granular media. 

Pulses in our experiments broaden, become smoother 
and rapidly attenuate as they travel away from the impact site.
%
Pulse propagation is not spherically symmetric about the site of impact. Pulse peak velocities are about twice as large along a direction directly down compared to those along a direction of about 45$^\circ$ from vertical.  This behavior may be 
similar to phenomena seen in simulations of oblique impacts that showed plastic deformation within the granular medium extending further laterally than vertically, and led to a more rapid decay of energy in the lateral pulse compared to the vertical pulse \citep{Miklavcic_2022}.


We measure a pulse duration of approximately 1 ms which most closely matches a seismic source
time equal to the crater size divided by the pulse travel speed.  This suggests that this time is relevant for pressure release within the material during crater formation.   To within a factor of a few,
the hypothesis that the seismic source time is similar to the pulse duration is supported
by the experimental measurements of pulse duration in higher velocity experiments by \citet{yasui15} and \citet{Matsue_2020}.
The pulses in sand were about half as long as those in millet, 
suggesting that pulse duration could also be sensitive to grain size.   Our pulse durations could also be consistent with an initially short 
duration pulse (less than 1 ms)
and subsequent broadening, but only if the diffusion coefficient (for propagating velocity or
pressure perturbations) 
is about an order of magnitude larger than expected from
the experiments of weak pulses into granular media by \citet{Langlois_2015}.

Our detected acceleration and velocity signals exhibit attenuation and broadening.
Pulse shapes become smoother as they propagate.  Because these are characteristics of diffusive 
behavior we adopted a diffusive rather than dispersive model for pulse propagation. 
Using a simple diffusive model for pulse broadening, we estimate
that the peak pulse velocity, pressure and seismic energy are proportional to $r^{-\frac{5}{2}}$ and
the peak acceleration $\propto r^{-3}$ where $r$ is distance from site of impact.    
The predicted power law index of -3 is consistent with decay of peak acceleration measured
in higher velocity experiments by \citet{Matsue_2020}.
Our measurements are roughly consistent
with these exponents for signals from accelerometers near the impact site where the pulses
are stronger and less affected by reflections from the container base or walls. 
These decay exponents are steeper than predicted or
measured in impacts into solids for pulse travel in the elastic regime \citep{Guldermeister_2017}.
The rapid attenuation in our experiments supports a seismic jolt model for impact
excited pulse propagation in rubble asteroids 
\citep{nolan92,greenberg94,greenberg96,Nolan_2001,thomas05} 
and does not support  the slowly attenuating seismic reverberation model \citep{cintala78,cheng02,richardson04,richardson05,Yamada_2016}.

Past about 5 ms after impact, our pulses can reach tub walls and base.   A positive pressure pulse
can reflect off a hard wall in a granular medium.  We do detect accelerations after 5 ms but they 
decay rapidly.  We have not tried to model the behavior of the accelerations at later
times because the signals decay rapidly and so are weak at later times, because reflections are complicated by flexure in the tub and because vibrational energy is absorbed through the tub base.
Nevertheless the rapid decay of seismic energy in the medium at later times
is not inconsistent with the rapid decay of energy at earlier times.  After 2 tub crossing times, signals are sufficiently weak that we cannot determine if there
are reflections off the granular surface.  Because
material can be ejected from a granular surface by a positive pressure pulse \citep{wright19}, we expect that seismic energy would be absorbed 
rather than reflected by the free granular surface. This could also be true in a low g environment such as a rubble asteroid
\citep{tancredi12}.

Using the kinetic energy in the pulse measured from the accelerometer nearest the impact site,
we estimate a seismic efficiency 
$k_{seis}$ of about a percent at a distance from impact site that is just outside the crater radius.
This seismic efficiency value is high compared to the higher velocity impact experiments into granular media by \citet{yasui15}. 
However, as attenuation is rapid, the estimated seismic efficiency is sensitive to the distance
at which seismic energy is measured.  Pulse strength might be
better characterized by the ratio of the momentum that
goes into the pressure pulse and that of the impactor (our $B_{\rm eff}$ parameter) than the seismic efficiency parameter ($k_{seis}$).


We apply our diffusive propagation model to estimate the size and exponents for decay
of pulse physical properties for the pulse excited by the forthcoming DART impact into Dimorphos, the secondary of the  asteroid Didymos.   We made the following assumptions:  
The asteroid Dimorphos is comprised of rubble. 
We adopt a pressure dependent pulse travel speed based on studies of granular systems \citep{vandenWildenberg_2013}. 
The initial pulse width
is set by the seismic source time at the crater radius.   
The pulse width subsequently broadens 
via a diffusive approximation with diffusion coefficient given by Equation \ref{eqn:D}. 
The radial component of momentum integrated at the pulse front per unit solid angle 
is conserved.  We assume that the seismic pulse propagates isotropically as a function of $r$.   By scaling from properties at a distance $r = R_{cr}$ equal to the crater radius from
the site of impact, we ignore the details of the early phases of the impact that would include vaporization, melting and strong shocks. 

The resulting models exhibit rapid attenuation of seismic energy due to pulse broadening. The models resemble a constant pulse velocity diffusive model which we compared to our experiments in Section \ref{sec:diff} and which have peak pulse pressure decaying as a function of travel distance from site of impact in power-law form $P_{pk} \propto r^{- \frac{5}{2}}$.
Using parameters estimated for the DART impact, and despite assuming a high diffusion coefficient, 
we estimate that pulse peak accelerations will exceed surface gravity as the pulse travels
through the asteroid. Because the surface of a granular system would have little cohesion,  we expect that the asteroid surface will absorb seismic energy by lofting and disturbing surface particles. We do not expect a reflected pulse, so pulse strength will be negligible after a single crossing of the asteroid and there will not be much reverberation. 
Material on the surface could be disturbed over a large fraction of the asteroid surface prior to the landing of crater ejecta.  The impact could be nearly catastrophic, in the sense that a significant layer of asteroid material would be lofted from the surface by the seismic pulse, though most of this material should return to re-accumulate on the core.  
Our experiments suggest that the pulse amplitude could be dependent on propagation angle, so some regions of the asteroid surface could be more strongly disturbed by the seismic pulse than others.

There is a discrepancy in our experiments between the strength of diffusion and those
estimated in the experiments by \citet{Langlois_2015}. While we took into account the pressure dependence of the pulse travel speed \citep{vandenWildenberg_2013} in our rough estimate for the properties of the impulse excited by the DART impact,  
we neglected a possible pressure dependence of the diffusion or attenuation rate.  A dependence of the broadening rate on pulse peak pressure might account for our high estimated diffusion coefficients compared to other experiments.  We estimate that, in our experiments, pulse peak pressures exceed hydrostatic pressure within about 10 cm of the impact site. Our pressure pulses are just barely in a regime that is relevant for a low-g environment.   
Perhaps the regime with peak pressure larger than hydrostatic pressure $P_{pk}>P_0$, which is relevant for low g environments, exhibits faster broadening than outside this regime, with peak pulse pressure below hydrostatic or confining pressure, $P_{pk} < P_0$.    
Future studies could better characterize the dependence of the attenuation or diffusion rate on the pulse amplitude. 
Our experiments suggest that the rate of pulse broadening and attenuation is sensitive to grain size.
Future work could study how grain properties (friction coefficients, elastic modulus) and shape, size and composition distributions affect propagation of impact generated pulses.  

Our experiments show that pulse amplitude depends on 
the direction of propagation and is higher normal to the surface (downward in our experiment). Future studies could improve measurements of the angular dependence of impact generated pulses and study simulations to learn how impact generated pulses depend on propagation angle from the surface normal, the impact angle and the surface slope.  


We adopted a diffusive empirical model to describe pulse broadening and attenuation in part because of its simplicity.  With a few assumptions and free parameters, pulse properties can be predicted as a function of travel distance from site of impact by scaling from values at the crater radius. If a diffusive model gives an approximate description of pulse propagation in granular media, it would be powerful because of its simplicity. 

In our model for the DART impact,  we assumed that pulse propagation in rubble in low g primarily is dependent on the peak pressure in the seismic pulse, following \citet{vandenWildenberg_2013}.   Dimensionless numbers indirectly enter into our pulse propagation model because we scaled the model from properties estimated at the crater radius which obeys crater scaling relations \citep{holsapple93,Housen_2003,yasui15}. 
If pulse propagation is primarily dependent on peak pulse pressure,  then our diffusive approximation can be directly applied to low g environments.
However, pulse propagation within granular systems is nonlinear and possibly dispersive as well as advective and diffusive, so future studies could test the validity of the approximations we have adopted here and their application to low g environments.

\vskip 2 truein
{Acknowledgements:}

This material is based upon work supported in part by NASA grant 80NSSC21K0143, and National Science Foundation Grant No. PHY-1757062.
This work was supported in part by an award to J. South from the Rochester Academy of Sciences for purchase of granular materials.  This work was supported in part by undergraduate research awards from the University of Rochester to J. South and N. Skerrett.   
We thank Anton Peshkov   
for helpful discussions and correspondence. 
We thank Jiaxin Wang for helping us in the lab. 
We thank Jack Mottley for advice on pulse generation techniques.

\if \ispreprint1
\subsection{Nomenclature}
\begin{table}
\caption{Nomenclature \label{tab:nomen}}
\begin{tabular}{lll}
\hline
Drop height & $h_{drop}$ \\
Projectile mass & $m_p$ \\
Projectile radius & $R_p$ \\
Projectile density & $\rho_p$  \\
Projectile velocity & $v_p$  \\
Projectile inverse stopping length & $\alpha_p$ \\
Impact velocity & $v_{imp}$ \\
Kinetic energy of impact  & $K_{imp}$ \\
Froude number & $Fr $ \\
Dimensionless numbers  & $\pi_2, \pi_R, \pi_4$ \\
Ambient or hydrostatic pressure  & $P_0$ \\
Peak pressure in a pulse & $P_{pk}$ \\
Peak velocity in a pulse & $v_{pk}$ \\
Peak acceleration in a pulse & $a_{pk}$ \\
Peal displacement in a pulse & $\delta_{pk}$ \\ 
Pulse propagation speed & $ v_P$ \\
Depth & $H$ \\
Distance from impact site & $r$ \\
Crater radius & $R_{cr}$ \\
Crater diameter & $D_{cr}$ \\
Gravitational acceleration on Earth & $g_\oplus, g$ \\
Surface gravitational acceleration & $g_{\rm eff}$ \\
Radial velocity component & $v_r$ \\
Radial acceleration component  & $a_r$ \\
Time of propagation & $ t_{prop},t_{a,prop}, t_{v,prop}$ \\
Pulse duration & $\Delta t, \Delta t_a, \Delta t_v $ \\
Normalized pulse duration & $W, W_a, W_v$ \\
Spatial pulse width & $\Delta r$ \\
Diffusion coefficient & $D$ \\
Diffusion scaling coefficient & $ C_W $ \\
Bulk density of substrate & $\rho_s$ \\
Grain diameter & $d_g$ \\
Grain elastic modulus  & $E_g$ \\
Sound speed in the grain's material & $c_g$ \\
Drag coefficient & $C_D$ \\
Seismic energy & $E_{seis}$ \\
Seismic efficiency & $ k_{seis} $\\
Momentum transfer parameters & $b_{\rm eff} , B_{\rm eff} $ \\
Asteroid mass & $m_a$ \\
Asteroid radius & $R_a$ \\
Angle of repose & $\theta_r$ \\
Coefficient of friction & $\mu_s$ \\
\hline
\end{tabular}
\end{table}
\fi

\bibliographystyle{elsarticle-harv}
\bibliography{refs_atten}

\end{document}